\documentclass[fleqn,usenatbib]{mnras}

\usepackage{newtxtext,newtxmath}

\usepackage[T1]{fontenc}

\DeclareRobustCommand{\VAN}[3]{#2}
\let\VANthebibliography\thebibliography
\def\thebibliography{\DeclareRobustCommand{\VAN}[3]{##3}\VANthebibliography}

\usepackage{graphicx}	
\usepackage{amsmath}	
\usepackage{amssymb}	
\usepackage{color,soul}

\usepackage{xcolor}

\usepackage{floatrow}
\usepackage{pdflscape}
\usepackage[labelfont=bf]{caption}

\title[MUSE-ALMA Halos VI]{MUSE-ALMA Halos VI: Coupling Atomic, Ionised \& Molecular Gas Kinematics of Galaxies}

\author[R. Szakacs et al.]{
Roland Szakacs,$^{1}$\thanks{E-mail: roland.szakacs@eso.org (ESO)}
Céline Péroux,$^{1,2}$
Martin Zwaan,$^{1}$
Aleksandra Hamanowicz,$^{3}$
\newauthor$\;$Anne Klitsch,$^{4}$
Alejandra Y. Fresco,$^{5}$
Ramona Augustin,$^{3}$
Andrew Biggs,$^{1}$
\newauthor $\;$Varsha Kulkarni,$^{6}$
Hadi Rahmani$^{2,7}$
\\
$^{1}$European Southern Observatory (ESO), Karl-Schwarzschild-Str. 2, 85748 Garching bei München, Germany\\
$^{2}$Aix Marseille Univ, CNRS, CNES, LAM, Marseille, France\\
$^{3}$Space Telescope Science Institute, 3700 San Martin Dr, Baltimore, MD 21218\\
$^{4}$DARK, Niels Bohr Institute, University of Copenhagen, Jagtvej 128, 2200 Copenhagen, Denmark\\
$^{5}$Max-Planck-Institut für Extraterrestrische Physik (MPE), Giessenbachstr. 1, 85748 Garching bei München, Germany\\
$^{6}$University of South Carolina, Dept. of Physics and Astronomy, 712 Main Street, Columbia, SC 29208, USA \\
$^{7}$GEPI, Observatoire de Paris, PSL Université, CNRS, 5 Place Jules Janssen, 92190 Meudon, France}

\date{Accepted XXX. Received YYY; in original form ZZZ}

\pubyear{2020}

\begin{document}
\label{firstpage}
\pagerange{\pageref{firstpage}--\pageref{lastpage}}
\maketitle

\begin{abstract}
We present results of MUSE-ALMA Halos, an ongoing study of the Circumgalactic Medium (CGM) of galaxies ($z \leq$ 1.4). Using multi-phase observations we probe the neutral, ionised and molecular gas in a sub-sample containing six absorbers and nine associated galaxies in the redshift range $z \sim 0.3-0.75$. Here, we give an in-depth analysis of the newly CO-detected galaxy Q2131-G1 ($z=0.42974$), while providing stringent mass and depletion time limits for the non-detected galaxies. Q2131-G1 is associated with an absorber with column densities of $\textrm{log}(N_\textrm{HI}/\textrm{cm}^{-2}) \sim 19.5$ and $\textrm{log}(N_{\textrm{H}_2}/\textrm{cm}^{-2}) \sim 16.5$, has a star formation rate of $\textrm{SFR} = 2.00 \pm 0.20 \; \textrm{M}_{\odot} \textrm{yr}^{-1}$, a dark matter fraction of {$f_\textrm{DM}(r_{1/2}) = 0.24 - 0.54$} and a molecular gas mass of {$M_\textrm{mol} = 3.52 ^{+3.95}_{-0.31} \times 10^9 \; \textrm{M}_{\odot}$} resulting in a depletion time of {$\tau_\textrm{dep} < 4.15 \; \textrm{Gyr}$}.  Kinematic modelling of both the CO (3--2) and [OIII] $\lambda 5008$ emission lines of Q2131-G1 shows that the molecular and ionised gas phases are well aligned directionally and that the maximum rotation velocities closely match. These two gas phases within the disk are strongly coupled. The metallicity, kinematics and orientation of the atomic and molecular gas traced by a two-component absorption feature is consistent with being part of the extended rotating disk with a well-separated additional component associated with infalling gas. {Compared to emission-selected samples, we find that HI-selected galaxies have high molecular gas masses given their low star formation rate. We consequently derive high depletion times for these objects.}

\end{abstract}




\section{Introduction}

One of the most puzzling questions in galaxy evolution is how galaxies sustain their star formation. Due to the short depletion timescales that have been observed, it is evident that galaxies have to accrete gas from an external source in order to maintain their continuity on the main sequence \citep[e.g.][]{Scoville_2017}. The inflowing gas is accreted from the intergalactic medium (IGM). While it is a challenging task to observe the accretion process due to the low density of the extragalactic gas, a number of inflows have been observed over the last few years \citep[e.g.][]{Rubin_2012, Martin_2012, Turner_2017, Zabl_2019}. Metal enriched gas is also expelled from galaxies due to AGN feedback \citep{Shull_2014} or stellar feedback \citep[e.g.][]{Ginolfi_2020}. A fraction of the expelled gas is returned through galactic fountains \citep[]{Fraternali_2017, Bish_2019}, where cooled down gas rains onto the galactic disk, while some other part is returned to the IGM through galactic winds driven by AGN and stellar feedback processes. 

The inflowing and outflowing gas interacts in a zone called the circumgalactic medium (CGM), which is loosely defined as the gas surrounding galaxies outside of the disk or ISM, but within the virial radius \citep{Tumlinson_2017}. While it can be a challenging endeavour to observe the CGM directly, due to the low surface brightness of the gas \citep[e.g.][]{Frank_2012, Corlies_2018, Augustin_2019}, observations and simulations indicate that the CGM is a multi-phase medium. The hot phase of the CGM has been observed through X-ray observations \cite[e.g.][]{Anderson_2010,Anderson_2013,Bregman_2018, Nicastro_2018} and Ly-$\alpha$ emission \cite[e.g.][]{Cantalupo_2014, Wisotzki_2016, Wisotzki_2018, Umehata_2019}. The cooler gas in the CGM can be probed by studying absorption lines in quasar (QSO) spectra, which offer the advantage of the sensitivity being independent of redshift \citep[e.g.][]{Tripp_1998}. This cooler low density gas has been detected through the absorption lines of various metal species and Hydrogen \citep[e.g.][]{Steidel_2010, Rudie_2012, Werk_2013, Turner_2014}. Hydrodynamic simulations strengthen the picture of a multi-phase CGM, by finding a mixture of cooler ($T \sim 10^4$ K) and hotter ($T \sim 10^{5.5} - 10^6$ K) gas within the virial radius of simulated galaxies \citep[e.g.][]{Stinson_2012, Suresh_2017, Nelson_2020}.

An important aspect to understand how galaxies sustain their star formation is to connect the CGM gas probed by absorption with the galaxies associated with the absorbers. {Narrow-band imaging and long-slit spectroscopic studies have searched for nebular emissions from HI-selected galaxies \mbox{\citep[e.g.][]{Kulkarni_2000, Kulkarni_2001}} and have, in part, been successful in the past \mbox{\citep[e.g.][]{Chen_2005, Fynbo_2010}}.} Further, integral field spectroscopy (IFS) combined with long slit spectroscopy follow ups have made it possible to not only associate galaxies with strong HI absorbers \citep[e.g.][]{Bouche_2007, Peroux_2011a, Peroux_2011b, Peroux_2017, Rudie_2017}, but to also study the star formation rate, metallicity of the emission line gas and kinematics of the ionised gas \citep[e.g.][]{Bouche_2012, Peroux_2017, Rahmani_2018a, Rahmani_2018b, Hamanowicz_2020}. The findings, among others, include a correlation between the SFR of the associated galaxy and the equivalent width of the absorption, indicating a physical connection between star bursts and gas seen in absorption \citep{Bouche_2007}. \cite{Peroux_2011a} find that in the majority of the cases the metallicity of the absorption is lower than of the associated galaxy. The number of studies associating absorption features found in the spectra of quasars with physical properties of absorber host candidates is low. An additional issue remains: associating galaxies with absorbers that are in complex group environments as studied in this and a previous MUSE-ALMA Halos publication \citep{Hamanowicz_2020}. The authors suggest that galaxies found in these environments would benefit from associating the kinematics of the galaxies with the absorber in order to distinguish which galaxies/environments the absorption is tracing \citep[e.g. see][]{Rahmani_2018a}.  Therefore obtaining more observations of absorber - absorber host systems plays a key part in furthering the understanding of the medium surrounding galaxies. 

Searches at the radio / sub-mm wavelengths with instruments like the Atacama Large Millimeter Array (ALMA), have enabled the community to study the mass, depletion time, and kinematics of the molecular gas in galaxies associated with absorbers \citep[e.g.][]{Neeleman_2016, Neeleman_2018, Klitsch_2018, Klitsch_2019, Augistin_2018,  Moller_2018, Kanekar_2018, Kanekar_2020, Peroux_2019, Freundlich_2021}. {One of the findings that these HI-selected galaxies have in common is that the molecular gas masses of these galaxies are high for their given SFR, leading to depletion times that are up to multiple factors larger than the averages found in emission selected galaxies. Further observations and constructing statistically significant samples, like the ones obtained in the MUSE-ALMA Halos project, are needed in order to study a possible correlation.}

Obtaining spatially-resolved multi-phase data of galaxies have furthered our understanding in how the ionised and molecular gas phases relate to each other. Kinematic studies have revealed that the two phases mostly align well spatially \citep[e.g.][]{Uebler_2018, Moller_2018, Klitsch_2018, Loiacono_2019, Peroux_2019, Molina_2019, Molina_2020}. Further kinematic studies by the EDGE-CALIFA survey \citep{Levy_2018} have shown that 75\% of the galaxies in their sample have higher maximum rotational velocities for the molecular gas while the remaining 25\% have similar maximum rotational velocities to the ionised gas. \cite{Peroux_2019} on the other hand did indeed find a case where the rotational velocity of the molecular gas was significantly lower than for the ionised gas in a galaxy associated with a strong HI-absorber. The number of galaxies observed in both the molecular and ionised gas phase is still low and studies of these gas phases is a key point in furthering our understanding of gas flows within and surrounding galaxies.

Another aspect of using IFS-based multi-wavelength observations is that these data make it possible to estimate the dark matter fractions in the inner parts of galaxies. A widely accepted notion is that dark matter dominates the outskirts of galaxies, however the distribution of matter in the central parts of galaxies is still debated. Studies like the DiskMass survey \citep{Martinsson_2013} have observed 30 spiral galaxies at the current epoch and found the central dark matter fractions to be mostly in the range of 0.5-0.9. Studies of higher redshift galaxies find lower central dark matter fractions in both observations and simulations \citep[e.g.][]{Uebler_2018, Uebler_2020a, Genzel_2017, Genzel_2020}. \cite{Price_2020} report a decrease of the dark matter fraction toward higher redshifts, attributed to various intertwined effects of galaxy mass–size growth, gas fraction, and halo growth and evolution. Therefore obtaining further samples of central dark matter fractions are an important aspect of understanding the reasons for the differences in the central dark matter fractions over different epochs.

The studied MUSE-ALMA Halos sub-sample includes six absorbers and nine associated galaxies in the redshift range $z \sim 0.3 - 0.75$. In this publication we present the results from new ALMA observations of the fields Q2131-1207, Q1232-0224, Q0152-2001, Q1211-1030, Q1130-1449 each of which contain a strong HI absorber at $z \sim$ 0.4 and in the case of Q1232-0224 an additional one at $z \sim$ 0.75. While we analyze and provide information on all fields, the focus of this publication lies on the CO-detected galaxy Q2131-G1 in the field Q2131-1207 {(first reported in \mbox{\cite{Bergeron_1986}} and further analyzed in \mbox{\cite{Guillemin_1997}} and \mbox{\cite{Kacprzak_2015}})}.

The paper is organized as follows: Section 2 presents the observational set-up and data reduction and imaging process. Section 3 describes the molecular properties of the galaxies associated with the strong HI absorbers, while describing both the physical and morpho-kinematical properties and providing limits for non detections. In section 4 we discuss our findings and put them into context with previous observations. Finally, section 5 gives a summary of the findings. Throughout this paper we adopt an $H_0 = 70 \textrm{km s}^{-1} \textrm{Mpc}^{-1}, \Omega_\textrm{M} = 0.3, \textrm{and} \; \Omega_\Lambda = 0.7$ cosmology.

\section{Observations}
\label{sec:obs}

We follow a multi-wavelength approach in order to study the gas and associated galaxies in this study, combining VLT/MUSE, HST and ALMA observations. The observations and corresponding data processing/imaging are presented in this section.

\subsection{Optical campaign}

\subsubsection{VLT/MUSE Observations}

In this work we study five fields containing quasar absorbers (Q2131-1207, Q1232-0224, Q0152-2001, Q1211+1030 and Q1130-1449). These fields are a subset of the full MUSE-ALMA Halos sample which have ALMA follow-up observations targeting {redshifts} of $z \sim$ 0.4 and $z \sim 0.75$. That sample has been observed using VLT/MUSE in period 96 under programme ESO 96.A-0303 (PI: C. Peroux). All fields were observed in nominal mode (4800-9400 \AA) under good seeing conditions (<0.85 arcsec). The first four fields were observed for 1-2 hours per target, while Q1130-1449 was observed significantly deeper ($12 \times 1200$s). The observations and data reduction method for the 5 quasar fields is described in depth in \mbox{\cite{Peroux_2019}} and \mbox{\cite{Hamanowicz_2020}}. In short, the ESO MUSE reduction pipeline v2.2 \mbox{\citep{Weilbacher_2016}} was used. Bias, flat and wavelength calibration was applied in addition to line spread functions and illumination correction frames to each individual exposure. These astrometry solutions and the correction for geometry and flux calibrations where then applied. Each of the individual exposures were  combined including field rotation. Instead of the pipeline sky subtraction method, the sky emission lines were removed using a Principal Component Analysis algorithm \mbox{\citep[][]{Husemann_2016}}. Additionally, the MUSE observations for the fields have been discussed in depth in the following publications: Q0152-020 \citep{Rahmani_2017,Rahmani_2018, Hamanowicz_2020}; Q1130-1449 \citep{Peroux_2019, Hamanowicz_2020}; Q2131-1207 \mbox{\citep{Peroux_2017, Hamanowicz_2020}}; Q1232-0224, Q1211-1030 \mbox{\citep{Hamanowicz_2020}}.

\subsubsection{HST Observations}

We select fields that show strong HI absorption column densities in quasar spectra. The column densities are based on literature and were derived using data from the Faint Object Spectrograph (FOS) and Cosmic Origins Spectrograph (COS) on HST (see \mbox{\cite{Boisse_1998}} for details about the $\textrm{Q2131z039}_\textrm{HI}$ absorber; \mbox{\cite{Lane_1998}} for $\textrm{Q1130z031}_\textrm{HI}$; \mbox{\cite{Rao_2006}} for $\textrm{Q1232z075}_\textrm{MgII}$; \mbox{\cite{Muzahid_2016}} for $\textrm{Q2131z043}_\textrm{HI}$ and $\textrm{1211z039}_\textrm{HI}$; \mbox{\cite{Rahmani_2018b}} for $\textrm{Q0152z038}_\textrm{HI}$).

Readily available and reduced archival HST imaging is used for observations of the stellar continuum. The exposure times for the five fields range from 10 to 50 minutes. Observations of Q2131-1207 (PI: Maccheto, ID:5143) , Q1232-0224 (PI: Bergeron, ID:5351), Q0152-2001 (PI: Steidel, ID:6557) and Q1211+1030 (PI: Bergeron, ID:5351) use the Wide Field Planetary Camera 2 (WPFC2) in the F702W filter. The observation of Q1130-1449 (PI: Bielby, ID: 14594) uses the Wide Field Camera 3 (WFC3) in filter IR-F140W.

Further archival HST data, obtained with the Cosmic Origins Spectrograph (COS) on HST, are used for studying the $\textrm{H}_2$ absorption lines of the absorber associated with Q2131-G1. Specifically, we use these spectra to study the position of the $\textrm{H}_{2}$ absorption line in velocity space. The $\textrm{H}_{2}$ absorption has been extensively studied in \mbox{\cite{Muzahid_2016}}. We use two observations with a wavelength range of 1140-1800\AA$\:$, which consist of G130M (exposure time: 77 minutes) and G160M (exposure time: 120 minutes) FUV grating integrations at a medium resolution of $R \sim$ 20,000 (corresponding to a Full Width Half Maximum (FWHM) of $\sim$ 18 km s$^{-1}$).  (PI: Churchill, ID: 13398). Due to the Lyman-limit break of the absorber ($z = 0.43$) there is no recorded QSO flux at wavelengths below 1310\AA.

Additionally we have an ongoing HST multi-band photometry program of 40 orbits (PI: Péroux, ID: 15939). This program will allow us to study the morphology and stellar masses of galaxies associated with HI absorbers in the MUSE-ALMA Halos survey.

This program will allow us to study the morphology and stellar masses of 200 $z$ < 1.2 galaxies associated with HI and MgII absorbers (including our current sample) in more detail at a later stage of the MUSE-ALMA halos project. 

\subsection{ALMA Observations}
\subsubsection{Observation Details}
The fields Q2131-1207, Q1232-0224, Q0152-2001 and Q1211+1030 were observed with ALMA in Band 6 in order to cover the CO(3--2) lines of galaxies associated with absorbers found at $z\sim0.4$ (programme 2017.1.00571.S, PI: C. Peroux). Given the field of view (FOV) of ALMA in band 6 we target a subset of galaxies previously observed with MUSE with impact parameters ranging from 8 to 82 kpc. All of the fields have one spectral window that was centred on the redshifted CO(3--2) frequency of 345.796 GHz with a high spectral resolution mode. This results in 3840 channels, each with a 1.129 MHz width. Additional three other spectral windows are also included for these observations in a low spectral resolution mode (31.250 MHz). The CO(3--2) line of one the galaxies in the field Q1232-0224 ($z$ = 0.7566) is expected to be in one of the low resolution spectral windows. We also include the previously studied field Q1130-1449 in our analysis. Details concerning this observation can be found in \cite{Peroux_2019}.

A table with the quasar coordinates, observation dates, exposure times, angular resolution, used calibrators, percipitable water vapour (PWV) and antenna configurations for the different observed fields can be found in Table \ref{tbl:quasar_spec} in the appendix (section \ref{sec:alma_obs}).

\subsubsection{Data Reduction and Imaging}

In this section we describe the image processing of the fields Q2131-1207, Q1232-0224, Q0152-2001 and Q1211-1449 observed with ALMA. The fields are imaged and, when possible, self-calibrated using the Common Astronomy Software Applications package \citep[\textsc{CASA},][]{McMullin_2007} version 5.6.2-3. 

As a starting point for all imaging and calibration the pipeline-calibrated $uv$-datasets as delivered by ALMA-ARC are used. When multiple measurement sets (MS) are provided due to multiple observations, we combine them using the \texttt{concat} task. Using these combined measurement sets we reconstruct an initial continuum image of the field by using the task \texttt{tclean}. Depending on the synthesized beam size, we use different pixel sizes for the imaging (0.18" for Q0152-2001, 0.17" for Q1211-1030, 0.2" for Q2131-G1 and 0.22" for Q1232-0224). For all datasets we use \texttt{tclean} with a Briggs weighting scheme with the robust parameter set to 1.0, a standard gridder and a hogbom deconvolver. 

In the case of Q0152-2001 and Q1211-1030 we follow up \texttt{tclean} with the task \texttt{uvcontsub} in order to subtract the central quasar in the field. As a final step we use the continuum-subtracted uv-dataset and the task \texttt{tclean} with the same parameters as for the continuum images and a spectral binning of 50 km $\textrm{s}^{-1}$. 

Both the quasars in Q2131-1207 and Q1232-0224 are bright at mm-wavelengths, allowing us to perform self-calibration. Therefore, after creating the initial model and continuum image mentioned above, we calculate the temporal gains using the task \texttt{gaincal} with gaintype \texttt{G} (which determines the gains for each polarization and spectral window) using a solution interval of 35 s for Q2131-1207 and 70 s for Q1232-0224. For both calibrations we check that the solutions show a smooth evolution over time and that the solutions have an acceptable signal-to-noise ratio (SNR) > 10. Then we apply the solutions to the measurement sets using the task \texttt{applycal} in linear interpolation mode and create an updated sky model and continuum image using \texttt{tclean}. Following the phase calibration we proceeded with a second round of amplitude calibration using gaintype G and a solution interval of 105 s for Q2131-1207 and 70 s for Q1232-0224. Following this we create another updated sky model and continuum image using \texttt{tclean}. Then we follow up with the continuum subtraction using \texttt{uvcontsub} with order 3 for Q2131-1207 and 2 for Q1232-0224. We use the continuum-subtracted dataset to create a data cube using \texttt{tclean} with the same parameters as for the continuum images and a spectral binning of 50 km $\textrm{s}^{-1}$. As the final step we produce a cube corrected for the primary beam using the \texttt{impbcor} task. The final RMS for the cubes where self-calibration was feasible is $\sim 1.5 \times 10^{-4} \; \textrm{Jy}$. The cubes where no self-calibration was possible have an RMS $\sim 2.8 \times 10^{-4} \; \textrm{Jy}$.

\begin{landscape}
\begin{table}
 \begin{tabular}{||c c c c c c c c c c c||} 
 \hline
 \textbf{Absorber ID} &
 $\textrm{\textbf{z}}_\textrm{\textbf{abs}}$ ${}^\textrm{\textbf{a}}$ &
 $\textrm{\textbf{log(N}}_\textrm{\textbf{HI,abs}})$ &
 $[\textrm{\textbf{Fe/H}}]_{\textrm{\textbf{abs}}} $ \\
 
 & 
 & 
 [$\textrm{cm}^{-2}$] & \\
 
 \hline
 Galaxy | ($\textrm{z}$) &
 b ${}^\textrm{\textbf{a}}$ &
 $\textrm{SFR}_\textrm{[OII]}$ ${}^\textrm{\textbf{a}}$ &
 $12+\textrm{log(O/H)}_{\textrm{l}}$ ${}^\textrm{\textbf{a}}$ &
 $12+\textrm{log(O/H)}_{\textrm{u}}$ ${}^\textrm{\textbf{a}}$ &
 $\textrm{f}_\textrm{CO}$ &
 $\textrm{S}_{\textrm{CO}}$ &
 $\textrm{FWHM}_\textrm{CO} $ & 
 $\textrm{L}_{\textrm{CO(1-0)}} $ &
 $\textrm{M}_\textrm{mol}$ &
 $\tau_\textrm{dep}$ \\
 
 &
 [kpc / '']&
 $[\textrm{M}_{\odot}\textrm{yr}^{-1}$] &
 &
 &
 [GHz] &
 [Jy $\textrm{km} \: \textrm{s}^{-1}$]&
 [km $\textrm{s}^{-1}$] &
 [$10^9$ K km $\textrm{s}^{-1}$ $\textrm{pc}^{2}$] &
 [$10^9 \textrm{M}_{\odot}$]&
 [Gyr] \\
 \hline\hline
$\textrm{\textbf{Q2131z043}}_{\textrm{\textbf{HI}}}$  & \textbf{0.43} & \textbf{19.5} $\mathbf{\pm}$ \textbf{0.15} ${}^\textrm{\textbf{b}}$ & $\mathbf{> -0.96} $ ${}^\textrm{\textbf{a}}$\\
\hline
Q2131-G1 | (0.42974) & 52 / 9.2 & $2.00 \pm 0.2$ & $8.98 \pm 0.02$ & - & 241.866 & $0.36 \pm 0.02$ & $184 \pm 50$ & $1.42 \pm 0.08$ & {$3.52 ^{+3.95}_{-0.31}$} & {< 4.15} \\
 \hline
Q2131-G2 | ($0.4307 \;^\textrm{\textbf{a}}$) & 61 / 10.7 & $0.20 \pm 0.1$ & $8.32 \pm 0.16$ & - & 241.697 & < 0.068 & - & < 0.27 & {< 3.64} & {< 36.37} \\
 \hline
 
$\textrm{\textbf{Q1232z039}}_{\textrm{\textbf{HI}}}$  &
\textbf{0.3950} &
\textbf{20.75} $\mathbf{\pm}$ \textbf{0.07} ${}^\textrm{\textbf{c}}$ &
$\mathbf{< -1.31} $ ${}^\textrm{\textbf{c}}$\\
\hline
Q1232-G1 | ($0.3953\;^\textrm{\textbf{a}}$) & 8 / 1.5 & $0.67 \pm 0.09$ & $8.02 \pm 0.06 $ & $8.66 \pm 0.04 $ & 247.829 & {< 0.070} & - & < 0.24 & {< 6.09} &  {< 8.02}\\
 \hline
 
$\textrm{\textbf{Q1232z075}}_{\textrm{\textbf{MgII}}}$  &
\textbf{0.7572} &
$\mathbf{18.36}^\mathbf{+0.09}_\mathbf{-0.08}$ ${}^\textrm{\textbf{d}}$ &
$\mathbf{> -1.48} $ ${}^\textrm{\textbf{d}}$\\
\hline
Q1232-G2 | ($0.7566\;^\textrm{\textbf{a}}$)& 68 / 9.1 & $2.58 \pm 0.23$ & $8.19 \pm 0.19 $ &  $8.54 \pm 0.19 $& 262.462 & < 0.12 & - & < 0.83 & {< 18.31} & {< 7.80}\\
 \hline
 
$\textrm{\textbf{Q0152z038}}_{\textrm{\textbf{HI}}}$  & \textbf{0.3887} & $\textrm{<} $\textbf{18.8} ${}^\textrm{\textbf{e}}$ & $\mathbf{> -1.36} $ ${}^\textrm{\textbf{a}}$\\
\hline
Q0152-G1 | ($0.3826\;^\textrm{\textbf{a}}$) & 60 / 11.5 & $1.04 \pm 0.03$ & $8.65 \pm 0.09 $ & - & 250.105 & < 0.17 & - & < 0.53 & {< 2.80} & {< 2.78}\\
 \hline
$\textrm{\textbf{Q1211z039}}_{\textrm{\textbf{HI}}}$  & \textbf{0.3929} & \textbf{19.46} $\mathbf{\pm}$ \textbf{0.08} ${}^\textrm{\textbf{b}}$ & $\mathbf{> -1.05} $ ${}^\textrm{\textbf{a}}$\\
\hline
Q1211-G1 | ($0.3928\;^\textrm{\textbf{a}}$)& 37 / 6.8 & $4.71 \pm 0.08$ & $8.16 \pm 0.01 $ & $8.48 \pm 0.01 $ & 248.274 & < 0.15 & - & < 0.49 & {< 6.78} & {< 1.47}\\
 \hline
 $\textrm{\textbf{Q1130z031}}_{\textrm{\textbf{HI}}}$  & \textbf{0.3127} & \textbf{21.71} $\mathbf{\pm}$ \textbf{0.07} ${}^\textrm{\textbf{f}}$ & \textbf{-1.94} $\mathbf{\pm}$ \textbf{0.08} ${}^\textrm{\textbf{g}}$  \footnotemark[1]\\
\hline
Q1130-G2 | ($0.3127\;^\textrm{\textbf{a}}$)& 44 / 9.5 &
$0.44 \pm 0.3$ &
$8.77 \pm 0.05 $ &
- &
263.4 &
0.63 $\pm$ 0.01 ${}^\textrm{\textbf{g}}$ &
$250 \pm 50$ ${}^\textrm{\textbf{g}}$  &
3.1 $\pm$ 0.1 ${}^\textrm{\textbf{g}}$ &
$11.03^{+1.44}_{-1.27}$ &
25$^{+21}_{-20}$ \\
\hline
Q1130-G4 | ($0.3126\;^\textrm{\textbf{a}}$)&
82 / 17.7 &
$> 0.40 $ &
$< 8.65$ &
- &
263.44 &
0.42 $\pm$ 0.03 ${}^\textrm{\textbf{g}}$& 
$535 \pm 50$ ${}^\textrm{\textbf{g}}$  &
2.1 $\pm$ 0.1 ${}^\textrm{\textbf{g}}$ &
{> 8.88} &
{$\gtrless 22.19$} \\
 \hline
Q1130-G6 | ($0.3115\;^\textrm{\textbf{a}}$)&
98 / 21.3 &
$1.14 \pm 0.7$ &
$8.94 \pm 0.16$ &
- &
263.67 &
$0.20 \pm 0.01$ ${}^\textrm{\textbf{g}}$ &
$205 \pm 50$ ${}^\textrm{\textbf{g}}$ &
$1.0 \pm 0.1$ ${}^\textrm{\textbf{g}}$ &
$2.65^{+1.20}_{-0.82}$ &
$2.3^{+1.4}_{-1.1} $\\
\hline
\end{tabular}
\caption{\label{tbl:gal_abs_spec} \textbf{Physical properties of absorption-selected galaxies.} \newline 
Row 1 - absorber: (1) reference name of the absorber used in this paper, (2) redshift of the absorber, (3) HI column density of the absorber, (4) metallicity of the absorber \newline
Row 2 - galaxy: (1) reference name of the galaxy used in this paper and its redshift in brackets, (2) impact parameter in kpc and arcseconds, (3) star formation rate measured from the [O II] emission line (not dust corrected), (4/5) lower/upper metallicity 12+log(O/H) (not dust corrected, both metallicity branches derived by \protect\cite{Hamanowicz_2020} are displayed following \protect\cite[][]{Kobulnicky_1999}). If only one metallicity branch is reported in literature, the upper column is left blank. (6) observed CO flux density, (7) CO velocity width, (8) CO(1-0) Luminosity, (9) molecular mass, (10) depletion timescale of the galaxy. \newline
Literature references: ${}^\textrm{\textbf{a}}$) \protect\cite{Hamanowicz_2020} , ${}^\textrm{\textbf{b}}$) \protect\cite{Muzahid_2016}, ${}^\textrm{\textbf{c}}$) \protect\cite{Boisse_1998}, ${}^\textrm{\textbf{d}}$) \protect\cite{Rao_2006}, ${}^\textrm{\textbf{e}}$) \protect\cite{Rahmani_2018b}, ${}^\textrm{\textbf{f}}$) \protect\cite{Lane_1998},
${}^\textrm{\textbf{g}}$) \protect\cite{Peroux_2019}.}
\end{table}
\renewcommand*{\thefootnote}{\arabic{footnote}} \footnotetext[1]{We note that {\cite{Kanekar_2009}} reports a higher metallicity of $\textrm{[Z/H]}_\textrm{abs} = -0.90 \pm 0.11 $ for this absorber.}
\end{landscape}

\section{MOLECULAR GAS PROPERTIES OF THE GALAXIES
ASSOCIATED WITH THE ABSORBERS}

We target nine galaxies in the redshift range $z = 0.31-0.76$. Out of those nine galaxies we detect four: the previously detected galaxies Q1130-G2, Q1130-G4 and Q1130-G6 \citep[presented in][]{Peroux_2019} and the newly {CO-detected} galaxy Q2131-G1. We provide an analysis of the physical and morpho-kinematical properties of Q2131-G1 in this section. Additionally, we provide stringent limits on the molecular gas content of undetected galaxies. All the calculated physical properties of the targeted galaxies can be found in Table \ref{tbl:gal_abs_spec} and the morpho-kinematical properties are listed in Table \ref{tbl:morpho_kin_prop}.

\subsection{Properties of the CO-detected Galaxy (Q2131-G1, $z$ = 0.42974)}

In this section we describe the physical and morpho-kinematical properties of the CO-detected galaxy Q2131-G1 and the galaxy-gas (absorber) connection.

\subsubsection{Molecular Gas Mass and Depletion Time}
\label{subsec:phyiscal properties}

We study the molecular gas properties of the CO-detected galaxy Q2131-G1. We create an integrated flux map using the \textsc{CASA} task \texttt{immoments} and set the threshold of pixels to be counted above $\sim$2 $\sigma$ of the created cube. This integrated flux map yields an observed CO(3--2) flux of $S_{\textrm{CO}} = (0.36 \pm 0.02) \; \textrm{Jy km s}^{-1}$. We derive the CO(1-0) luminosity by first calculating ${L'}_\textrm{CO(3--2)}$ using $\textrm{S}_{\textrm{CO}}$ and the prescription by \cite{Solomon_1992}. Then we use the ${L'}_\textrm{CO(3--2)}$ to ${L'}_\textrm{CO(1-0)}$ conversion factor from \cite{Fixsen_1999}: ${L'}_\textrm{CO(3--2)} / {L'}_\textrm{CO(1-0)} = 0.27$ and obtain a CO(1-0) luminosity of ${L'}_{\textrm{CO(1-0)}} = (1.42 \pm 0.08) \times 10^9 \textrm{K km s}^{-1} \textrm{pc}^2$. We choose the Milky Way spectral line energy distribution conversion factor due to the rather low redshift of the galaxy ($z$=0.42974). We note that absorption-selected systems may preferentially select interacting galaxies, which have more excited CO SLEDs than isolated galaxies making the used SLED a first order approximation for Q2131-G1 \citep[][]{Klitsch_2019b}. The molecular mass is calculated by using the geometric mean of the \cite{Bolatto_2013} and \cite{Genzel_2012} $\alpha_\textrm{CO}(\textrm{Z})$ prescription. This conversion factor is a good approximation for galaxies which are not significantly below solar metallicity and therefore appropriate for Q2131-G1 (12+log(O/H) = $8.98 \pm 0.02$ \mbox{\citep{Peroux_2017}}, also see \cite{Genzel_2015} for a more detailed description of this averaged conversion factor). We note that \mbox{\cite{Muzahid_2016}} derived a lower metallicity, closer to the solar metallicity, for Q2131-G1 (12+log(O/H) = $8.68\pm0.09$). This discrepancy can be explained by the use of the N2-index, which is known to saturate at solar metallicities \mbox{\citep{Pettini_2004}}. We elect to use the $\textrm{R}_{23}$ based metallicity by \mbox{\cite{Peroux_2017}}, but note that the metallicity is based on emission line fluxes that have not been dust-corrected and therefore possibly overestimate the metallicity. We therefore base the conversion factor on 12+log(O/H) = $8.98 \pm 0.02$, but include the lower metallicity in the error calculation {and compute $\alpha_\textrm{CO} = 2.48 ^{+2.50}_{-0.08} \; \textrm{M}_{\odot} \textrm{(K km/s pc)}^{-1}$}. The molecular mass is  {$M_\textrm{mol} = 3.52 ^{+3.95}_{-0.31} \times 10^9 \textrm{M}_\odot$}. The calculated molecular mass is consistent with the mass limit of $M_\textrm{mol} \leq 7 \times 10^9 \textrm{M}_\odot$ using $L'_\textrm{CO(2-1)} \leq 3.8 \times 10^{9} \textrm{K km s}^{-1} \textrm{pc}^{2}$ \citep[][]{Klitsch_2020}.

Using the non-dust corrected star formation rate (SFR) derived by \cite{Hamanowicz_2020} ($\textrm{SFR}_\textrm{[OII]} = 2.00 \pm 0.2 \textrm{M}_\odot \textrm{yr}^{-1}$) we calculate the limit on the depletion time using:

\begin{equation}
\tau_{\textrm{dep}} < \frac{\textrm{M}_\textrm{mol,max}}{\textrm{SFR}_{\textrm{[OII,min]}}} \textrm{yr}
\label{eq:t_dep}
\end{equation} 

\noindent The depletion time for Q2131-G1 is {$\tau_{\textrm{dep}} < 4.15 \; \textrm{Gyr}$}.

\subsubsection{Stellar Mass}

In this section we estimate the stellar mass of Q2131-G1. The stellar mass is derived from the Mass-Metallicty-Relation (MZR) \citep{Tremonti_2004}. This relation is based on $\sim53000$ galaxies at $z \sim 0.1$ from the Sloan Digital Sky Survey (SDSS) sample and holds for $8.5 < \textrm{log(}M_\star / \textrm{M}_\odot) < 11.5$. Using the metallicity derived by \mbox{\cite{Peroux_2017}} (12+log(O/H) = $8.98 \pm 0.02$) we get two solutions: $\textrm{log(}M_\star / \textrm{M}_\odot \textrm{)} = 10.1 \pm 0.1$ and $\textrm{log(}M_\star / \textrm{M}_\odot \textrm{)} = 12.9 \pm 0.1$. This relation does not hold for the second solution, as that stellar mass would be outside of the valid range. We attempt to break this degeneracy by applying the Tully-Fisher relation (linking the stellar mass with the maximum rotation velocity of the galaxy) \citep{Tully_1977}. We used the relation by \cite{Puech_2008}, derived from a sample of $z \sim$ 0.6 galaxies using kinematics from the [OII] line. Using $V_\textrm{max} = 200 \pm 3 \: \textrm{km s}^{-1}$ (as derived by the kinematical analysis of the [OIII] $\lambda5008$ line in \cite{Peroux_2017}) we estimate the stellar mass of G2131-G1 to be $\textrm{log(}M_\star / \textrm{M}_\odot \textrm{)} = 10.54 \pm 0.71 $. This stellar mass is consistent with the lower stellar mass derived from the MZR. For further calculations we decide to use the stellar mass derived from the MZR, but take into account the value derived by the Tully-Fisher relation and by the MZR using the \mbox{\citep{Muzahid_2016}} metallicity (12+log(O/H) = $8.68 \pm 0.09$, $\textrm{log(}M_\star / \textrm{M}_\odot \textrm{)} = 9.1^{+0.3}_{-0.2}$) in the error calculations: $\textrm{log(}M_\star / \textrm{M}_\odot \textrm{)} = 10.1^{+0.5}_{-1.0}$

\subsubsection{Dark Matter Fraction}
\label{subsubsec:dm_frac}

\begin{figure*}
  \includegraphics[width=0.6\textwidth]{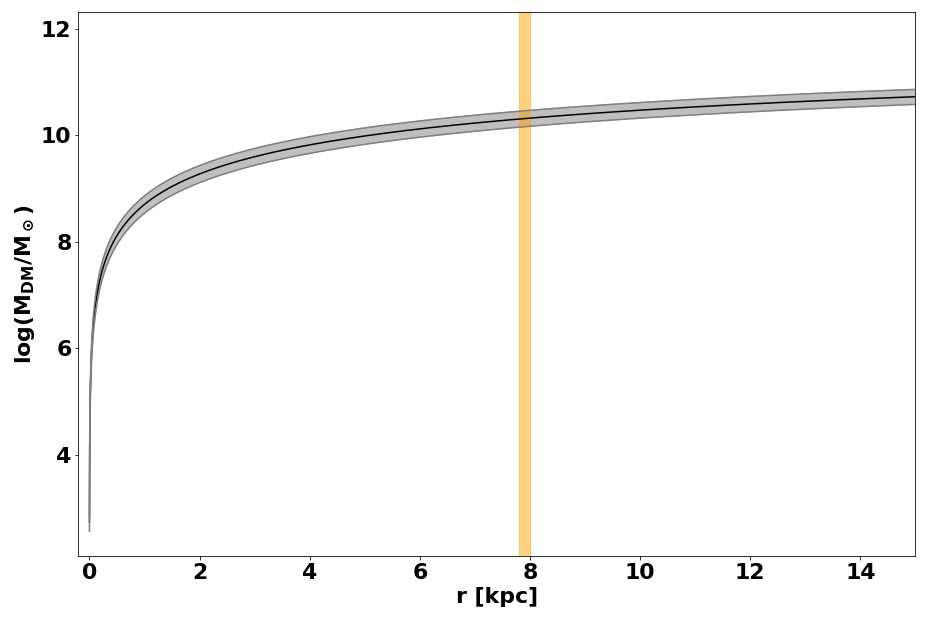}
  \caption{Cumulative mass of the dark matter within Q2131-G1 derived assuming a NFW profile. The shaded regions show the profile for the minimum/maximum derived dark matter mass $\textrm{M}_{200}$. The vertical orange line marks the [OIII] $\lambda5008$ half light radius. The dark matter fraction within the half-light radius is in the range of {$f_\textrm{DM} = 0.24-0.53$}. Therefore, we find the central regions of this galaxy to be baryon dominated.} 
  \label{fig:Q2131G1_dm_mass_prof}
\end{figure*}

Current studies have shown a declining dark matter fraction with increasing redshift \citep[e.g.][]{Genzel_2020, Price_2020}. We constrain the  dark matter contribution to the galaxy within the half-light radius. We create a NFW-profile \citep{Navarro_1997} based on the halo mass estimate (see section \ref{subsec:phyiscal properties}) and compute the corresponding cumulative mass curve. We note that this is a first order approximation of the dark matter fraction within the central region of G2131-G1.

The halo mass estimate is based on abundance matching (e.g \cite{Behroozi_2010}, \cite{Moster_2010} and \cite{Moster_2018}). We use the prescription provided in \cite{Genzel_2020} (equation A13 in \mbox{\cite{Genzel_2020}}, provided in a priv. comm. with B. Moster) based on the galaxy halo pairs from \cite{Moster_2018} to fit a halo mass - galaxy mass relation. This relation is appropriate for $z$ > 0.5 and provides an estimate of the halo mass derived from the stellar mass. Using the stellar mass of $\textrm{log(}M_\star / \textrm{M}_\odot \textrm{)} = 10.1^{+0.5}_{-1.0}$ we compute a halo mass of log($M_\textrm{200} / \textrm{M}_\odot) = 11.6 \pm 0.5$. This halo mass is consistent with the halo mass derived by \cite{Peroux_2017} assuming a spherical virialized collapse model by \cite{Mo_2002} {(log($M_\textrm{200} / \textrm{M}_\odot) = 12.46^{+0.03}_{-0.04}$)}. {The corresponding radius ($r_{200}$), within which the mean mass density is $\sim 200$ times the critical density of the Universe, is calculated using:}

\begin{equation}
    \label{equ:r_200}
    {r_{200} = \left[ \frac{M_{200}}{\frac{4}{3} \; \pi \; 200 \; \rho_{crit}}  \right]^{\frac{1}{3}}} \; \; ,
\end{equation}

\noindent {with $\rho_{crit}$ being:}

\begin{equation}
    \label{equ:rho_crit}
    \rho_\textrm{crit} = \frac{3 \; H^2(z)}{8 \; \pi \; G} \; \; ,
\end{equation}

\noindent {and using:}

\begin{equation}
    \label{equ:H_z}
    {H(z) = H_0 \sqrt{\Omega_M \; (1+z)^3 + \Omega_\Lambda} \; \;.}
\end{equation}

\noindent {Using equations \mbox{\ref{equ:r_200}}, \mbox{\ref{equ:rho_crit}} and \mbox{\ref{equ:H_z}} we compute: $H(z) = 87.9 \; \textrm{km s}^{-1} \textrm{Mpc}^{-1}$, $\rho_\textrm{crit} = 437.61 \; h^2 \; \textrm{M}_{\odot} \textrm{kpc}^{-3}$ and $r_{200} = 133^{+54}_{-36} \; \textrm{kpc}$.}

In order to fully describe the {NFW mass-profile} we compute the concentration parameter ($c$) which we compute {using the redshift dependent NFW concentration-mass relation from \mbox{\cite{Dutton_2014}}}:

\begin{equation}
    {\textrm{log}(c) = a + b \times \textrm{log}(M_{200} / [10^{12} h^{-1} \textrm{M}_\odot]) \; \; ,}
\end{equation}

\noindent {with}:

\begin{gather}
    {a = 0.520 + (0.905 - 0.520) \times \textrm{exp}(-0.617 \times z^{1.21})} \\
    b = -0.101 + 0.026 \times z \; \; ,
\end{gather}

\noindent {and with $\delta_{c}$ being:}

\begin{equation}
    \delta_c = \frac{200}{3} \times \frac{c^3}{\textrm{ln}(1+c) - \frac{c}{1+c}} \; \; .
\end{equation}

\noindent Using our derived $M_\textrm{200}$ we find {$c = 7.5^{+0.7}_{-0.7}$ and $\delta_{c} = 22430^{+4924}_{-4364}$.}

\noindent We calculate the {NFW mass-profile} using:

\begin{equation}
    M_\textrm{DM}(r) = 4 \pi \rho_0 r^3_\textrm{s} \times \left[\textrm{ln}\left(1+\frac{r}{r_\textrm{s}}\right) - \frac{\frac{r}{r_\textrm{s}}}{1+\frac{r}{r_\textrm{s}}}\right] \; \; .
\end{equation}

\noindent with $\rho_0 = \delta_\textrm{c} \rho_\textrm{crit}$ and $r_\textrm{s} = \frac{r_{200}}{c}$.

\noindent The resulting mass profile is shown in Fig. \ref{fig:Q2131G1_dm_mass_prof}. The dark matter mass is in the range of {$\textrm{log}(M_\textrm{DM}(r_{1/2})/\textrm{ M}_\odot) = 10.16 - 10.46 $} at the {[OIII] $\lambda5008$ emission half-light} radius $r_{1/2} = 7.9 \pm 0.1$ kpc. 

{We calculate the dynamical mass within $r_{1/2}$ using \mbox{\citep[][]{Epinat_2009}}:}

\begin{equation}
    M_\textrm{dyn}(r_{1/2}) = \frac{V^2(r_{1/2}) \; r_{1/2}}{G} \; \; ,
\end{equation}

\noindent {with $V(r_{1/2})$ being computed using an arctan velocity profile with the fit parameters derived by $\textrm{GalPak}^{\textrm{3D}}$ using the [OIII] $\lambda 5008$ emission line \mbox{\citep[][]{Bouche_2015}}}:

\begin{equation}
    {V(r_{1/2}) =  V_\textrm{max} \; \frac{2}{\pi} \; \textrm{arctan}(\frac{r_{1/2}}{r_\textrm{t}}) \; \; ,}
\end{equation}

\noindent {with $r_\textrm{t} = 1.51 \; \textrm{kpc}$ being the turnover radius. The velocity at the half-light radius is therefore {$V(r_{1/2}) = 176 \pm 3 \; \textrm{km s}^{-1}$} and the dynamical mass at {$r_{1/2}$ is $M_\textrm{dyn}(r_{1/2}) = 10.75 \pm 0.3$}. Using the dynamical mass and the dark matter mass within the half-light we compute the dark matter fraction within the half-light radius to be {$f_\textrm{DM} = 0.24 - 0.54$.}}

\begin{figure*}
  \includegraphics[width=\textwidth]{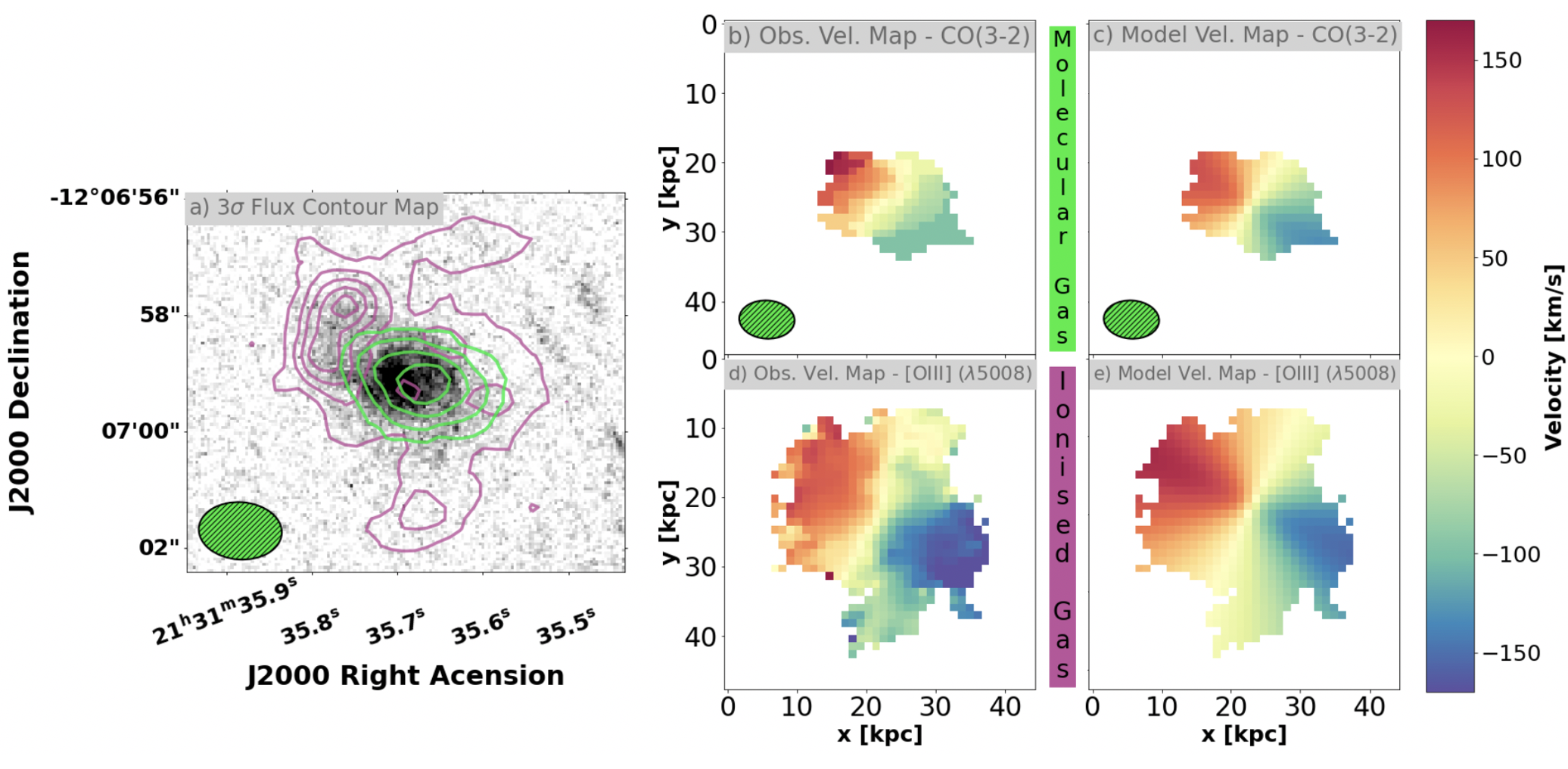}
  \caption{Contour plot and velocity maps of Q2131-G1. Ionised gas contour plot and velocity maps are based on [OIII] $\lambda5008$. Molecular gas contour plot and velocity maps are based on CO (3--2). a) {HST image of Q2131-G1 \citep[detector: PC, filter: F702W,][]{Kacprzak_2015} overlayed with} contour plots of the [OIII] $\lambda5008$ (purple) and CO(3--2) flux using 3$\sigma$ steps. b) Observed velocity map of the molecular gas c) Model velocity map of the molecular gas d) Observed velocity map of the ionised gas. e) Model velocity map of the ionised gas. In the contour plot a region of high [OIII] $\lambda5008$ flux is visible where no CO (3--2) is being observed {above the $3 \sigma$ threshold}. The direction of rotation for both the ionised and molecular gas are closely correlated and both gas phases show similar maximum rotational velocities.}
  \label{fig:Q2131G1_cont_vel}
\end{figure*}

\subsubsection{Morphological and Kinematical Properties}
\label{subsec:morph_kine}

We first study the morphological properties of Q2131-G1 based on the flux observed with HST, MUSE and ALMA. The HST image {(filter: F702W)} of Q2131-G1 with overlayed contours from the observed [OIII] $\lambda$5008 and CO(3--2) observed flux maps can be found in Fig. \ref{fig:Q2131G1_cont_vel} a). CO(3--2) has a compact and {elliptical} morphology in the centre of the galaxy with an extent $\sim 20$ kpc. We stress that these {higher-z} observations would not resolve small-scale clumps as observed in the PHANGS-ALMA survey \citep[][]{Schinnerer_2019}. The ionised gas shows a greater extent of  $\sim 40$ kpc and has a shape that indicates spiral arms or possible tidal tails \citep{Peroux_2017}. There is a region of a [OIII] $\lambda5008$ flux maximum {(at $\sim 21^\textrm{h} 31^\textrm{m} 35.77^\textrm{s}, -12^{\circ} 06\textrm{'} 57.8\textrm{''}$)} where CO(3--2) is not detected above the $3 \sigma$ threshold. This region coincides with a spiral structure in Q2131-G1 and therefore most likely to a region of active star formation. A large fraction of the molecular gas in this region is possibly already depleted due to the star formation process, leading to a CO flux density below the {$3\sigma$ threshold}. The stellar continuum observed by HST extends beyond the molecular gas emission {above the $3\sigma$ threshold}.

We study the kinematics of both the ionised and molecular gas of the detected galaxy using the 3D fitting algorithm $\textrm{GalPak}^{\textrm{3D}}$ \citep{Bouche_2015}. The algorithm assumes a disk parametric model with 10 free (but also optionally fixable) parameters and probes the parameter space by implementing a Monte Carlo Markov Chain approach with non-traditional sampling laws. The algorithm provides stable results if the signal-to-noise (SNR) per spaxel of the brightest spaxel in the cube is SNR>3. Additionally, the half-light radius has to satisfy the  condition $r_{1/2}$/FWHM > 0.75 in order for the algorithm to converge, with the FWHM being the Full Width Half Maximum of the Point Spread Function (PSF). The $r_{1/2}$/FWHM ratio of Q2131-G1 is below that condition $r_{1/2}$/FWHM $\sim 0.5$, but the algorithm nonetheless fully converges as we assessed from the MCMC chain. In order to be consistent with the ionised gas kinematic model used in \cite[][]{Peroux_2017} we also use the exponential flux profile and an arctan velocity profile as assumptions for the disk model. {We also ensured that the ALMA cube is in the same reference frame as the MUSE cube (BARY).} We additionally create two models with an exponential and tanh velocity profile, which yield different results, in order to take the differences in models into account for the error calculation of the derived properties. {The observations are well reproduced by a rotating disk, as can be assessed from the low residuals in the flux (Fig. \mbox{\ref{fig:2131G1_residual}}) and velocity residual maps (Fig. \mbox{\ref{fig:2131G1_vel_res}}) in the appendix.}

The morpho-kinematical properties of the ionised gas of Q2131-G1 derived from the [O III] $\lambda$5008 and H$\beta$ line in the MUSE observations are described in \cite{Peroux_2017}. The authors report the following: The maximum circular velocity is well constrained at $V_\textrm{max} = 200 \pm 3 \; \textrm{km s}^{-1}$ , the half light radius is found to be $r_{1/2} = 7.9 \pm 0.1 \; \textrm{kpc}$, the derived position angle is PA = $65 \pm 1 ^{\circ}$ and the inclination is $i_\textrm{CO} = 60.5 \pm 1.2$. Based on the derived flux, velocity and dispersion maps, \cite{Peroux_2017} argue that the galaxy is a large rotating disc, with a velocity gradient along the major axis and a dispersion peak at the center of the galaxy. Using this approach we create velocity maps of both the ionised and molecular gas. The observable [b) - ALMA CO(3--2), (d) - MUSE [OIII] $\lambda5008$)] and model [c) - ALMA CO(3--2), (e) - MUSE $\lambda5008$] velocity maps are shown in Fig. \ref{fig:Q2131G1_cont_vel}. We find that the rotational velocities for both the ionised and molecular gas are closely correlated. This is also the case for the model maximum velocities of both components ($V_\textrm{max,[OIII]} = 200 \pm 3 \; \textrm{km s}^{-1}$ and {$V_\textrm{max,CO} = 195^{+4}_{-30} \; \textrm{km s}^{-1}$}). {Both of the model velocities are consistent with the observed velocities of both components ($V_\textrm{max-obs,[OIII]} \sim 205 \; \textrm{km s}^{-1}$ and $V_\textrm{max-obs,CO} \sim 190 \; \textrm{km s}^{-1}$)}.

The derived inclination of the molecular and ionised gas in Q2131-G1 are {$i_\textrm{[OIII]} = 60.5 \pm 1.2 ^{\circ}$} and {$i_\textrm{CO} = 47^{+10^{\circ}}_{-1}$}. The position angles (PA) are {$\textrm{PA}_\textrm{[OIII]} = 65 \pm 1 ^{\circ}$} and $\textrm{PA}_\textrm{CO} = 59 \pm 2 ^{\circ}$. We conclude that the gas phases in Q2131-G1 are aligned directionally. 

{While the two models converge in terms of morpho-kinematical properties, they differ in redshifts (CO (3--2): $z_{\textrm{CO}}  = 0.42974 \pm 0.00001$, [OIII] $\lambda5008$: $z_{\textrm{[OIII]}}=0.42914 \pm 0.00001$, H$\beta$: $z_{\textrm{H}\beta} = 0.42950 \pm 0.00001$). The other [OIII] line in the spectrum is too weak and the [OII] line is disregarded due to its doublet nature.  We attribute this discrepancy to a combination of the wavelength calibration uncertainty of MUSE, which translates to a velocity uncertainty of $\sim 25 \; \textrm{km s}^{-1}$, and an underestimate of the errors provided by $\textrm{GalPak}^{\textrm{3D}}$. The ALMA frequency accuracy is set by the system electronics and is much better than the corresponding channel width of the cube ($50 \; \textrm{km s}^{-1}$). We therefore use the redshift derived from the CO (3--2) model as a zero-point in the analysis of the absorber and gas kinematics. We include the value of $z_{\textrm{H}\beta}$ and other uncertainties mentioned above to estimate an error of $\pm 100 \; \textrm{km s}^{-1}$ {($\sim 25 \; \textrm{km s}^{-1}$ MUSE velocity uncertainty + $\sim 75 \; \textrm{km s}^{-1}$ kinematical modelling uncertainty)} for the kinematic zero-point of the [OIII] emission line in the following study of the absorber and gas kinematics. {For the CO (3--2) zero-point we estimate an error of $\sim 75 \textrm{km s}^{-1}$ (kinematical modelling uncertainty).}}

\begin{table*}

 \begin{tabular}{||c c c c c c c c||} 
 \hline
 \textbf{Galaxy} &
 $r_{1/2, \textrm{[OIII]}}$ &
 $i_{\textrm{[OIII]}}$ &
 $\textrm{PA}_{\textrm{[OIII]}}$ &
 $V_\textrm{max,[OIII]}$ &
 $\textrm{log(}M_\textrm{dyn,[OIII]}$) &
 $\textrm{log(}M_\textrm{h,[OIII]}$) \\ [0.5ex] 
 
 \hline
 
 &
 $r_{1/2, \textrm{CO}}$ &
 $i_{\textrm{CO}}$ &
 $\textrm{PA}_{\textrm{CO}} $ &
 $V_\textrm{max,CO}$ \\ [0.5ex] 
 
 &
 [kpc]&
 [deg] &
 [deg] &
 [km $\textrm{s}^{-1}$] &
 [$\textrm{M}_{\odot}$] &
 [$\textrm{M}_{\odot}$]\\
  \hline \hline
  
 \textbf{Q2131-G1} &
 $7.9 \pm 0.1$ &
 {$60.5 \pm 1.2 $} &
 $65 \pm 1$ &
 $200 \pm 3 $ &
 $10.87 \pm 0.03$ &
 $11.7 \pm 0.1$ \\
 \hline
 
 &
 {$3.7^{+0.5}_{-0.1}$ }&
 {$47^{+10}_{-1}$} &
 $59 \pm 2$ &
 {$195^{+4}_{-30} $} &
 - &
 - \\
 \hline
 
 \textbf{Q1130-G2} &
 $14 \pm 2$ &
 $77 \pm 2$ &
 $131 \pm 2$ &
 $264 \pm 14 $ &
 $11.3 \pm 0.2$ &
 $12.9 \pm 0.1$ \\
 \hline
 
 &
 $2 \pm 1$ &
 $76 \pm 3$ &
 $117 \pm 2$ &
 $134 \pm 14 $ &
 - &
 -  \\
 \hline
 
 \textbf{Q1130-G4} &
 $9 \pm 2$ &
 $54 \pm 2$ &
 $86 \pm 2$ &
 $231 \pm 12 $ &
 $11.1 \pm 0.2$ &
 $12.7 \pm 0.1$ \\
 \hline
 
 &
 $6 \pm 1$ &
 $82 \pm 4$ &
 $84 \pm 2$ &
 $290 \pm 19 $&
 - &
 -  \\
 \hline

\end{tabular}
\caption{\label{tbl:morpho_kin_prop} \textbf{Morpho-kinematic properties of galaxies detected in both [OIII] and CO(1-0) / CO(3--2).} \newline
Row 1 - properties derived from [OIII]: (1) reference name of the galaxy used in this paper, (2) half-light radius, (3) inclination, (4) position angle, (5) maximum velocity, (6) dynamical mass, (7) halo mass. \newline
Row 2 - properties derived from CO(1-0) / CO(3--2): (1) -, (2) half-light radius, (3) inclination, (4) position angle, (5) maximum velocity, (6) dynamical mass, (7) halo mass. \newline
Literature references: The values for Q2131-G1 [OIII] are taken from \protect\cite{Peroux_2017} and the values for Q1130-G2 [OIII]/CO, Q1130-G4 O III/CO are taken from \protect\cite{Peroux_2019}}.
\end{table*}

\subsubsection{Galaxy - Gas Connection}
\label{subsubsec:gal_abs_conn}

Kinematical studies of the gas {in} the galaxies seen in emission and probed by the quasar sightlines allow us to probe what galaxy/environment the absorbing gas is tracing. We use an approach based on the model rotation curve obtained by $\textrm{GalPak}^{\textrm{3D}}$ to tackle this question.

\begin{figure*}
\includegraphics[width=\textwidth]{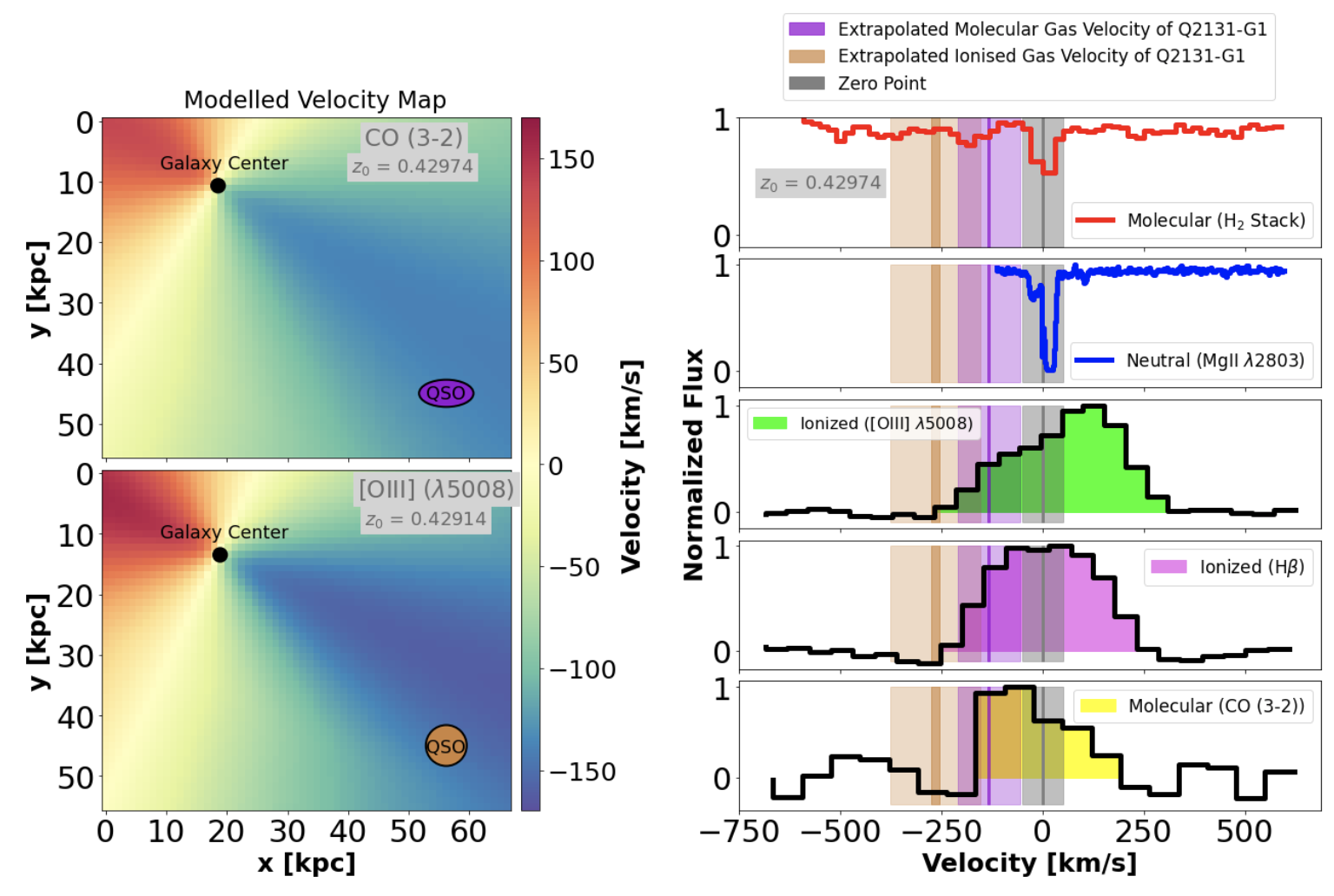}
\captionof{figure}{Extrapolated model velocity maps and normalized flux of the MgII ($\lambda2803$) and stacked $\textrm{H}_2$ absorption line and [OIII] $\lambda5008$, H$\beta$, CO(3--2) emission lines. The velocity zero point of the spectra is set to {the redshift of CO (3--2) derived from the kinematic study ($z_0 = 0.42974$) (displayed as the gray shaded area). The magenta and brown bars display the extrapolated velocities of the molecular and ionised gas of Q2131-G1 respectively. The shaded magenta and brown bars display the errors of the extrapolated velocities ({$75 \; \textrm{km s}^{-1}$} for the molecular gas, $100 \; \textrm{km s}^{-1}$ for the ionised gas). Extrapolating the model velocity maps derived from $\textrm{GalPak}^{\textrm{3D}}$ to the line of sight toward the quasar shows that at the position of the quasar the {molecular} and ionised gas of Q2131-G1 are located between $\sim -130 \; \textrm{and} -135 \; \textrm{km s}^{-1}$ and $\sim -255 \; \textrm{and} -275 \; \textrm{km s}^{-1}$ respectively, while the absorption features are found between $\sim -60 \; \textrm{and} +60 \; \textrm{km s}^{-1}$. We thus conclude that the two-component absorption features are consistent with in part an extended rotating disk of Q2131-G1 and in part gas falling onto Q2131-G1.}}
\label{fig:Q2131G1_spectra}
\end{figure*}

We extrapolate the rotation curves of Q2131-G1 for both MUSE and ALMA data to the line-of-sight (LOS) towards the quasar to relate it to the gas traced by the $\rm{H}_2$ and MgII absorber. The corresponding plots can be found in Fig. \ref{fig:Q2131G1_spectra} where we additionally show the normalized absorption and emission lines with the zero-point of velocity at {the redshift of CO (3--2) derived by $\textrm{GalPak}^{\textrm{3D}}$ ($z_0 = 0.42974$)}. {We find the extrapolated velocities of the molecular and ionised of Q2131-G1 between $\sim -130 \; \textrm{and} -135 \; \textrm{km s}^{-1}$ and $\sim -255 \; \textrm{and} -275 \; \textrm{km s}^{-1}$.} The absorption features, with column densities of $\textrm{log(}N_\textrm{HI} / \textrm{cm}^{-2}) = 19.5 \pm 0.15$ and $\textrm{log(}N_{\textrm{H}_2} / \textrm{cm}^{-2}) = 16.36 \pm 0.08$ {\mbox{\citep[][]{Muzahid_2016}}} {are found between $\sim -60 \; \textrm{and} +60 \; \textrm{km s}^{-1}$ {from the zero-point}}.

A limit on the CO absorption column density of the absorber $\textrm{Q2131z043}_\textrm{HI}$ associated with the galaxy Q2131-G1 is calculated following \citep[][]{Mangum_2015}, using an excitation temperature equal to the CMB temperature at the redshift, a $5 \sigma$ level from {the} spectrum {at the expected position and frequency of the CO(3--2) absorption line} as the detection threshold and a FWHM of 40 $\textrm{km s}^{-1}$ and derive $\textrm{log(}N_\textrm{CO} / \textrm{cm}^{-2}) < 14.6$. Using the mean ratio of $N_\textrm{CO} / N_{\textrm{H}_2} = 3 \times 10^{-6}$ \citep[][]{Burgh_2007} we derive $\textrm{log}(N_{\textrm{H}_2}/\textrm{cm}^{-2}) < 20.1$. This limit is consistent with the value observed from UV wavelength absorption by \cite{Muzahid_2016}.

{Studies of the absorption and emission metallicity connect the absorber to its host.} Using a metallicity gradient based on a sample of galaxy-absorber pairs {\mbox{\citep[$-0.022 \pm 0.004$ dex/kpc,][]{Christensen_2014}}}, we extrapolate the metallicity of Q2131-G1 to the LOS towards the quasar. {We take into account the observed flattening of the Oxygen metallicity gradient beyond $2 \times r_{1/2}$ \mbox{\citep{Sanchez_2016}} and assume that there is no change in the metallicity of the galaxy between $2 \times r_{1/2} = 15.8 \pm 0.2 \; \textrm{kpc}$ and the impact parameter $b = 52 \; \textrm{kpc}$}. {We use 12+log(O/H) = $8.98 \pm 0.02$ by \mbox{\cite{Peroux_2017}} as the metallicity of the galaxy, including the value by \mbox{\cite{Muzahid_2016}} (12+log(O/H) = $8.68 \pm 0.09$) in the error calculation.} The extrapolated metallicity of Q2131-G1 at the impact parameter ($b$ = 52 kpc) is {{$Z_\textrm{em} = - 0.06^{+0.09}_{-0.62}$}}. We additionally use an alternative metallicity gradient of $0.1/r_\textrm{1/2}$ {(which in the case of Q2131-G1 translates to $0.01266 \pm 0.00016 \; \textrm{dex/kpc}$)} derived by the CALIFA survey \citep[][]{Sanchez_2014} and find the extrapolated metallicity of Q2131-G1 at the impact parameter to be {$Z_\textrm{em} = 0.09^{+0.02}_{-0.48}$}. {Literature provides metallicity measurements using various species: $\textrm{[Fe/H]}_\textrm{abs} > -0.96$ from \mbox{\cite{Hamanowicz_2020}}, $\textrm{[O/H]}_\textrm{abs} = -0.26 \pm 0.19$ using ionisation modelling from \mbox{\cite{Muzahid_2016}}, the ionisation corrected metallicity of $\textrm{[S/H]}_\textrm{abs} > -0.72$  {(originally reported as $\textrm{[S/H]}_\textrm{abs} > -0.40$ assuming log($N_\textrm{HI,abs}/\textrm{cm}^{-2})$ = 19.18 instead of 19.5)} by \mbox{\cite{Som_2015}}). The global dust-free metallicity is $\textrm{[X/H]}_\textrm{abs} = -0.54 \pm 0.18$ \mbox{\citep{Peroux_2017}}.We find that both of the extrapolated metallicities are consistent with each other and consistent with the metallicity derived by \mbox{\cite{Peroux_2017}}}

\subsection{Limits from Non-Detections}
\label{subsec:limits}

{For the fields Q1232-0224, Q0152-2001 and Q1211-1030 with no CO detected counterparts in emission to the galaxies observed with MUSE and HST, we derive limits on the molecular mass and depletion times.}

For each cube we consider an ellipsoidal area with the minor axis, position angle and FWHM of the synthesized beam centered around the expected position of the galaxy with the frequency range being set to ± 100 km $s^{-1}$ centered around the redshifted frequency of the CO(3--2) emission line. We then assume the emission spectrum to be a Gaussian with an amplitude set to the RMS of the ellipsoidal area and a FWHM of 200 km $\textrm{s}^{-1}$. The flux limit is then the area under this line within the 5-$\sigma$ range.

We calculate the mass limits and depletion times following the same prescription as described in section \ref{subsec:phyiscal properties} (namely following \cite{Solomon_1992}, \cite{Fixsen_1999} and \cite{Genzel_2015}) and Equation \ref{eq:t_dep}. {The molecular gas mass limits use an $\alpha_{\rm{CO}}$ conversion factor based on the lowest measured metallicity of the galaxy to provide conservative limits of both the molecular gas mass and depletion time.}

The results for the CO flux, luminosity, mass and depletion time limits for Q0152-G1, Q1211-G1 and Q1232-G1 can be found in Table \ref{tbl:gal_abs_spec}. The CO(1--0) limits on the luminosity $\textrm{L}_\textrm{CO}$ are of the order $L_\textrm{CO} \sim 10^8 \; \textrm{K km} \textrm{s}^{-1} \textrm{pc}^2$, which fits the sensitivity estimates based on the ALMA sensitivity calculator calculated for our observations at $z \sim$ 0.4. {Our limits are more stringent than similar observations studying the molecular gas in objects associated with absorbers (e.g. MEGAFLOW by \mbox{\cite{Freundlich_2021}}, targeting galaxies around MgII absorbers, or \mbox{\cite{Kanekar_2018, Kanekar_2020}}), which are sensitive to luminosities $L_\textrm{CO} > \sim 10^9 \; \textrm{K km} \textrm{s}^{-1} \textrm{pc}^2$.} The molecular gas mass limits are in the range of {$\textrm{M}_\textrm{mol} \sim (2.8 - 18.3) \times 10^9 \; \textrm{M}_{\odot}$} and the depletion time limits are in the range of {$\tau_\textrm{dep} \sim 1.4 - 37 \; \textrm{Gyr}$}.

\section{Discussion}

The multi-wavelength approach in this work allows us to closely study the different gas phases within and around HI-selected galaxies. HST spectroscopy provides neutral and molecular gas information through absorption while MUSE and ALMA observations enable us to study the ionised and molecular gas content through emission. In this section we provide a detailed discussion of the observed properties and how they compare to current observations.

\subsection{Strongly Coupled Gas Phases within a Rotating Disk}

Recent observations of the ionised and molecular gas phases in galaxies between redshifts $z \sim 0.1 - 1.4$ have found that both phases mostly align well directionally \citep[e.g.][]{Uebler_2018, Moller_2018, Klitsch_2018, Loiacono_2019, Peroux_2019, Molina_2019, Molina_2020}. Similarly, we find that Q2131-G1 is well constrained by a disk model and that the ionised and molecular gas phase are aligned well directionally with similar inclinations and position angles.  

We also find a similar maximum rotational velocity ($\textrm{V}_\textrm{max}  \sim 200 \: \textrm{km s}^{-1}$) of the molecular and ionised gas within Q2131-G1. This is consistent with the EDGE-CALIFA survey \citep[][]{Levy_2018}, where ionised and molecular gas kinematics (traced by H$\alpha$) were compared in local galaxies. While the survey does find that for the majority of galaxies the rotational velocity measured from the molecular gas is higher than that from the ionised gas, there are cases {where} similar rotational velocities for both phases have been observed.

Due to the good alignment of the ionised and molecular gas phase, both directionally and rotationally, we find that the two gas phases are strongly coupled within Q2131-G1.

\subsection{Identifying the Disk Tilt}

\label{subsec:disk_tilt}

\begin{figure}
  \includegraphics[width=1.0\textwidth]{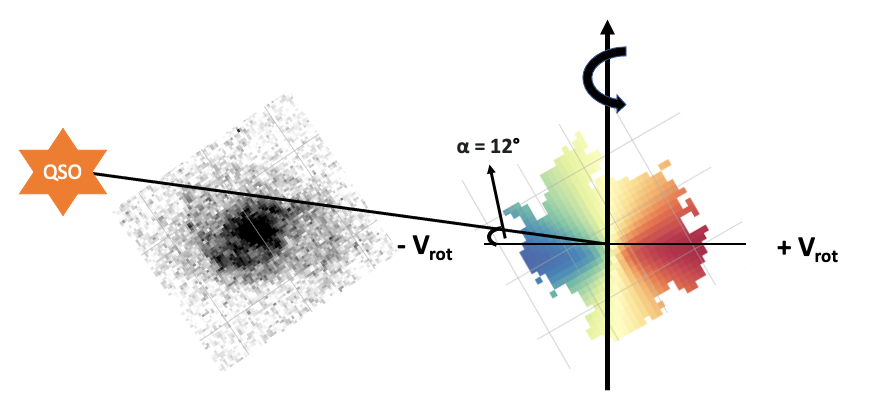}
  \caption{Sketch of the QSO - galaxy plane for identifying the disk tilt. {The galaxy is rotated in order to align the major axis with the x-axis in the sketch.} The spiral arms of Q2131-G1 wind in the opposite direction of the galaxies rotation and we conclude that the inclination has a negative sign.}
  \label{fig:2131G1_orient}
\end{figure}

Kinematic modelling {provides} the inclinations of both gas phases, but these values are degenerate without knowing the tilt of the disk.  A proposed solution to breaking the degeneracy of the disk tilt is to use the rotation curve and the winding direction of spiral arms \citep{Martin_2019}. Based on the likely assumption that in a self-gravitating, collisionless system only trailing spiral patterns are long lived \citep{Carlberg_1985} and most spiral patterns therefore lag behind the direction of rotation with increasing radius{, depending} on the winding rotation{,} one can infer a positive or negative sign of the inclination. The winding rotation of the spiral arms in Q2131-G1 observed in the HST image are opposite to the direction of rotation of the galaxy and the inclination therefore has a negative sign (see Fig. \ref{fig:2131G1_orient}).

\subsection{Gas Probed in Absorption Connected to a Rotating Disk and Infalling Gas}

\label{subsec:abs_rot_disk}
 
Previous authors state that an individual absorber is sometimes associated with multiple galaxies \citep[][]{Hamanowicz_2020}. In particular, in the field Q2131-1207 four galaxies are found at the same redshift and physically close to the absorber, indicating that Q2131-G1, found at $b=52$ kpc, is part of a group environment. Kinematical studies of the gas phases and the absorption features help alleviate these ambiguities studying how the different components relate in velocity space \citep[see e.g.][]{Rahmani_2017}.

To relate the gas probed in absorption with the absorber host we extrapolate the model rotation curve towards the sightline of the quasar in section \ref{subsubsec:gal_abs_conn} (see Fig. \ref{fig:Q2131G1_spectra}). {We find that the velocity of the ionised and molecular gas of Q2131-G1 at the point of the quasar sightline are blueshifted compared to the systemic redshift. A two-component absorption is found between $\sim -60$ and $60 \; \textrm{km  s}^{-1}$. {Due to the low azimuthal angle ($12^{\circ} \pm 1^{\circ}$) of Q2131-G1 \mbox{\citep[][]{Peroux_2017}} and simulations indicating that outflowing gas preferentially leaves the galaxy in a conical shape along its minor axis \mbox{\citep{Brook_2011, Peroux_2020}}, we assume an outflow scenario to be unlikely for both absorption components.}

{The weaker component is rotating in the same direction as the galaxy at less negative velocities. Further,} the extrapolated metallicities of Q2131-G1 ({$Z_\textrm{em} = - 0.06^{+0.09}_{-0.62}$} and {$Z_\textrm{em} = 0.09^{+0.02}_{-0.48}$}, depending on the metallicity gradient used) at the LOS towards the quasar indicate a connection between the gas probed in absorption and emission as it is consistent with the absorber metallicity ($\textrm{[X/H]}_\textrm{abs} = -0.54 \pm 0.18$). The extrapolated velocities and metallicities of the galaxy and the weaker absorption component are therefore consistent with being part of an extended rotating disk.

{The stronger absorption component is redshifted compared to the systemic redshift. Gas rotating with the disk of the galaxy is expected to have blueshifted velocities, making the stronger absorption component inconsistent with being part of the extended rotating disk. Further, the low azimuthal makes it a likely inflow \mbox{\citep[e.g.][]{Bordoloi_2011, Stewart_2011, Shen_2012}}.} The metallicity difference between the Q2131-G1 and the absorber lies in the infalling section of the galaxy to gas metallicity versus azimuthal angle plot seen in \mbox{\cite{Peroux_2016}} (figure 8 in the publication). Based on the metallicity difference and the geometry and orientation arguments, the stronger component is consistent with being gas falling onto Q2131-G1. We note that current data does not exclude that the gas could potentially also be falling onto Q2131-G2. The $\textrm{H}_{2}$ column density of the absorber also poses the question if and how it is possible to have {a considerable molecular gas phase, with temperatures down to 10 K, in infalling gas.}}

{We thus conclude that the two-component absorption features are consistent with in part an extended rotating disk of Q2131-G1 and in part gas falling onto Q2131-G1.}

\subsection{Specifics of HI-selected Systems}

\label{subsec:spec_HI_Sel}

Previous studies of HI-selected systems have observed gas depletion times that are a few times longer than what is typically found in surveys of emission-selected galaxies \citep[see especially][]{Kanekar_2018}. This poses the question whether the HI-selection preferentially selects galaxies that have large gas reservoirs for their SFR. We compare the detected galaxy Q2131-G1 with two current emission-selected molecular gas surveys, namely {xCOLD GASS \mbox{\citep[e.g.][]{Saintonge_2017}}} and the PHIBSS 1 \& 2 surveys \cite[e.g.][]{Tacconi_2018} {of} galaxies at redshift $z$ < 1.1. We additionally contrast with previously published HI-selected galaxies where molecular masses, stellar masses and SFR have been measured. We use a metallicity-dependent $\alpha_\textrm{CO}$ conversion factor for the comparison sample if metallicity information is provided \citep[namely][]{Genzel_2015, Bolatto_2013, Papa_2012}. Otherwise we use {$\alpha_{\textrm{CO}} = 4.3 \textrm{M}_{\odot} \textrm{(K km/s pc)}^{-1}$} from \citep{Bolatto_2013} or in the case of \cite{Klitsch_2018}  {$\alpha_{\textrm{CO}} = 0.6 \textrm{M}_{\odot} \textrm{(K km/s pc)}^{-1}$} from \cite{Papa_2012} is used because there is evidence that this galaxy is {a} luminous infrared galaxy (LIRG). In the case of the MUSE-ALMA Halos sample the SFR is not dust corrected (with exception of {the} field Q1130-1449), therefore the SFR can be considered as a lower limit.

\begin{figure*}
  \includegraphics[width=1.0\textwidth]{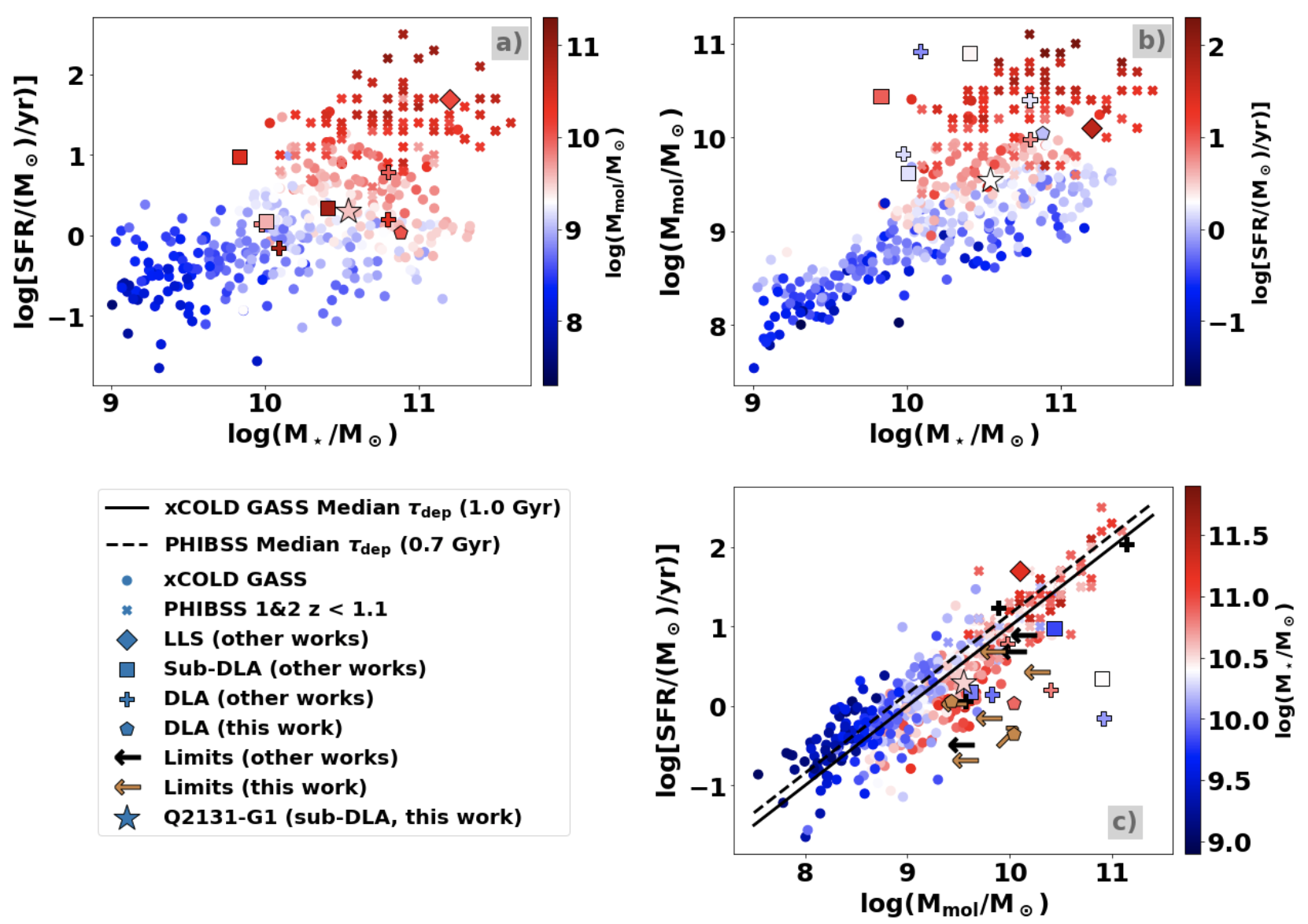}
  \caption{Star formation rate (SFR), molecular mass ($M_\textrm{mol}$) and depletion time ($\tau_\textrm{dep}$) plotted for HI-selected galaxies, the {xCOLD GASS} and PHIBSS 1\&2 survey (at $z$ < 1.1). HI-selected galaxies with molecular gas mass limits as well as galaxies without stellar mass data are additionally plotted in the SFR - $M_\textrm{mol}$ plot and have a {black (other works) / brown (this work) color}.  {The median depletion time for xCOLD GASS {\protect\citep[$\tau_\textrm{dep} = $ 1.0 Gyr,][]{Saintonge_2017}} and PHIBSS {\protect\citep[$\tau_\textrm{dep} = $ 0.7 Gyr,][]{Tacconi_2018}} are plotted as a black (dashed) line.} Q2131-G1 lies within the $M_\textrm{mol}$ - SFR, $M_{\star}$ -  $M_\textrm{mol}$ and $M_{\star}$ - SFR planes of the {xCOLD GASS} and PHIBSS 1\&2 surveys. We note that the SFR of Q2131-G1 is not dust corrected and therefore should be considered as a lower limit. {The molecular gas mass for the majority of sub-DLAs and DLAs for their given SFR is found to be higher than for emission-selected samples. This leads to depletion times in HI-selected galaxies that are up to multiple factors higher than for emission selected galaxies. This implies that selection based on strong HI-absorbers traces objects that have large gas reservoirs (at given SFR).} {Literature references: xCOLD GASS: {\protect\cite{Saintonge_2017}}; PHIBSS {\protect\cite{Tacconi_2018}}; LLS (other works): {\protect\cite{Klitsch_2018}}; sub-DLA (other works): {\protect\cite{Kanekar_2018, Neeleman_2016}}; DLA (other works): {\protect\cite{Kanekar_2018, Moller_2018, Neeleman_2018}}; Limits (other works): {\protect\cite{Klitsch_2018, Kanekar_2018}}.}}
  \label{fig:SFR_Mol_Star}
\end{figure*}

Fig. \ref{fig:SFR_Mol_Star} shows the SFR, molecular mass and depletion times of the emission and HI-selected galaxies. Q2131-G1 (star symbol) is comparable to the galaxies of the mass-selected {xCOLD GASS} and PHIBSS 1 \& 2 galaxies, as it lies within the $M_\textrm{mol}$ - SFR, $M_{\star}$ -  $M_\textrm{mol}$ and $M_{\star}$ - SFR planes. While Q2131-G1 fits well in the $\textrm{M}_\textrm{mol}$ - SFR plane, it is on the lower side of the derived SFR of comparable molecular masses, comparable to other galaxies associated with sub-DLAs. The deviations from the $\textrm{M}_\textrm{mol}$ - SFR plane are especially drastic in the case of galaxies associated with DLAs, which implies that HI-selection traces objects that have large gas reservoirs (at given SFR). Similarly the depletion timescale of Q2131-G1, Q1130-G2 and Q1130-G6 are an order of {$\sim 2 - 53$} larger than the median for emission-selected galaxies in the {xCOLD GASS} and PHIBSS  survey with $\tau_\textrm{dep,med} \approx 1.0$ Gyr and $\approx 0.7$ Gyr respectively. 

Studying this trend is limited due to the low number of molecular gas and star formation rate observations of HI-selected galaxies. Further studies will test whether HI-selection preferentially selects galaxies that have large molecular gas reservoirs for their given SFR.

\subsection{Connecting Galaxy Properties with Gas Properties}

\begin{figure*}
  \includegraphics[width=0.8\textwidth]{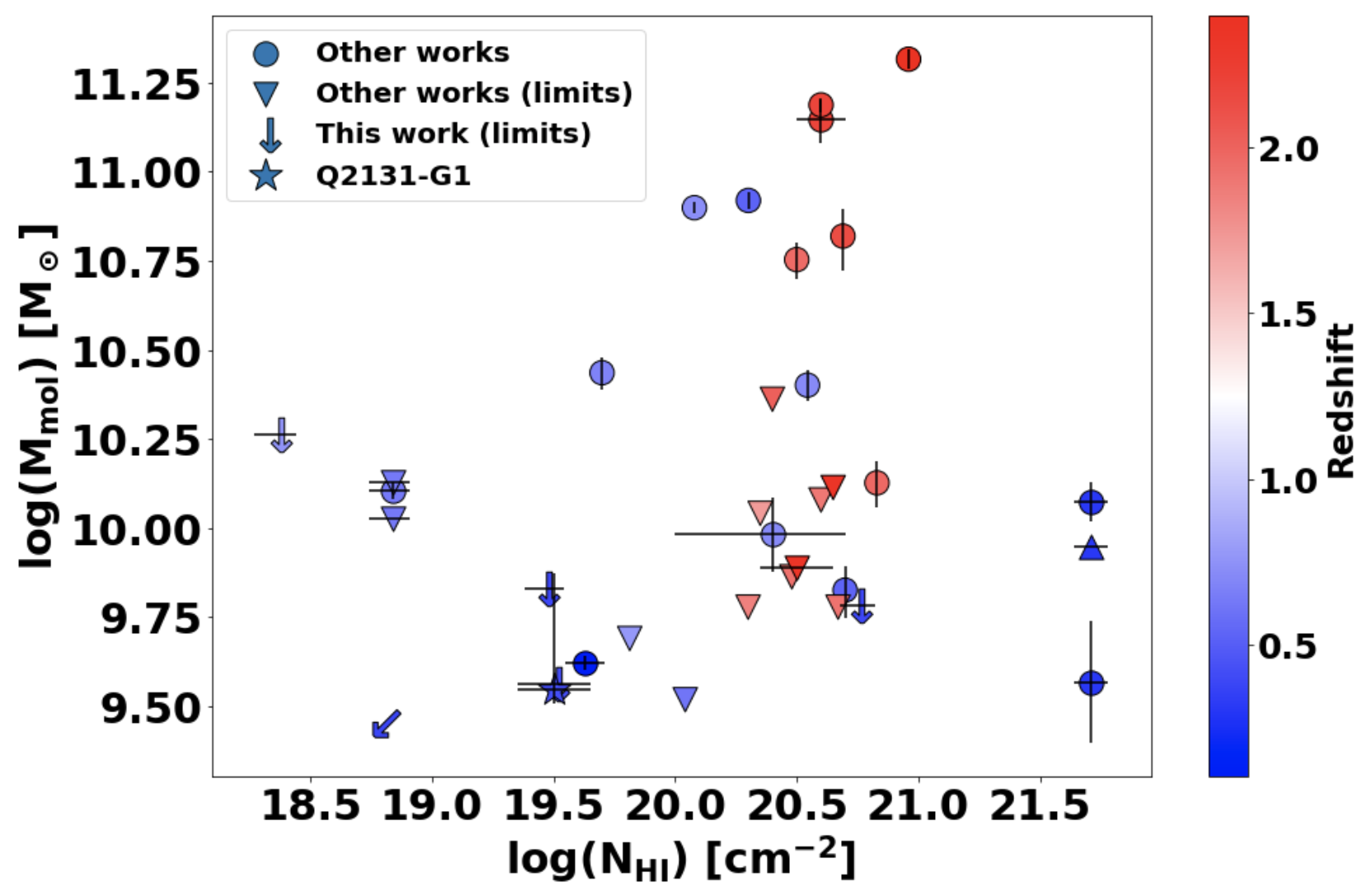}
  \caption{Absorber HI column density plotted against the molecular mass (limits) of absorber hosts by various published works \citep[][]{Neeleman_2016, Moller_2018, Augistin_2018, Kanekar_2018, Klitsch_2018, Neeleman_2018, Peroux_2019, Kanekar_2020} Q2131-G1 is on the lower side of previously detected molecular masses in HI absorption selected galaxies.}
  \label{fig:NHI_Mmol}
\end{figure*}

One key objective in studying absorption-selected galaxies is associating absorbers with potential absorber hosts and connecting absorber properties to the low density gas found by absorption.  We compare the derived molecular gas mass of Q2131-G1 and the HI column density of the associated absorber $\textrm{Q2131z043}_\textrm{HI}$ with {previously detected HI-absorbers and associated absorber hosts detected in CO} in Fig. \ref{fig:NHI_Mmol}. In order to provide a fair comparison, we use use the same conversion factors as described in section \ref{subsec:spec_HI_Sel}.

Molecular gas in HI-selected systems is found in systems with HI column densities between {$\textrm{log}(N_\textrm{HI}/\textrm{cm}^{-2}) \sim 18 - 22$}, from Lyman-limit systems \citep[][]{Klitsch_2018}, to sub-DLAs \citep[this work;][]{Neeleman_2016, Kanekar_2018} and DLAs \citep[][]{Moller_2018, Neeleman_2018, Kanekar_2018, Peroux_2019} (see Fig. \ref{fig:NHI_Mmol}). The molecular masses detected span over a large range of {$\textrm{log}(M_\textrm{mol}/\textrm{M}_{\odot}) \sim 9.5 - 11.3$}. The lower end of this range is typically for the detection limit of the observations. It is interesting to note that HI-selection can be associated with such large molecular gas reservoirs, but no correlation between the HI absorption column density and the absorber host molecular mass is seen. 

The most similar counterpart to Q2131-G1 is the galaxy associated with the absorber at redshift $z=0.101$ in the quasar spectrum of PKS 0439-433 \citep[][]{Neeleman_2016}. {While the absorber metallicity in PKS 0439-433 is higher \mbox{\citep[{[S/H]}= 0.1,][]{Som_2015}},} both absorbers show an HI-column density of $\textrm{log}(N_\textrm{HI}/\textrm{cm}^{-2}) \sim 19.5$ and the associated galaxies have closely matching molecular masses of $\textrm{log}(M_\textrm{mol}/\textrm{M}_{\odot}) \sim 9.6$. Additionally, both absorber systems have $\textrm{H}_{2}$ absorption features with $\textrm{H}_{2}$ column densities of $\textrm{log}(N_{\textrm{H}_2}/\textrm{cm}^{-2}) \sim 16.5$. The calculated limit on the CO column density ($\textrm{log}(N_{\textrm{CO}}/\textrm{cm}^{-2}) < 14.6$) in Section \ref{subsubsec:gal_abs_conn} and the subsequently derived limit on the H2 column density ($\textrm{log}(N_{\textrm{H}_2}/\textrm{cm}^{-2}) < 20.1$) is consistent with the detected $\textrm{H}_{2}$ column density. The galaxy in \cite{Neeleman_2016} does have a lower impact parameter of {$\sim 20$ kpc} than Q2131-G1 ($b = 52$ kpc), but the absorption features cannot be kinematically associated to the rotating disk of the absorber host {or infalling gas} and is likely part of the CGM of the galaxy. While we might probe different environments, the similarity of the molecular masses in the absorber hosts and the HI/$\textrm{H}_2$ column densities of the absorbers indicate a connection of these parameters. The other galaxies in this sample either lack observations of possible $\textrm{H}_2$ absorption features, or have not been detected at all. This is partly due to the low detection rates of $\textrm{H}_2$ in quasars ($\sim 16$ per cent for high-z absorbers \citep[][]{Noterdaeme_2008}, $\sim 50$ per cent for low-z absorbers \citep[][]{Muzahid_2015}). Nonetheless, future studies of molecular gas in both absorbers and absorber hosts, combined with kinematic studies that help to associate these systems are essential for studying a possible connection between the high density molecular gas found in galaxies and the low density molecular gas found in absorbers.

\subsection{A Dark Matter Fraction Evolving with Redshift}

Current extragalactic surveys of the dark matter fraction in the central regions of galaxies provide evidence for a dark matter fraction evolution with redshift, with the dark matter fraction declining for higher redshifts \citep[e.g.][]{Genzel_2020, Price_2020}. A possible explanation for this evolution of the dark matter fraction over different redshifts is given by the IllustrisTNG (TNG) simulations \citep[][]{Lovell_2018}. The authors find that the evolution is due to the more centrally concentrated baryonic mass at higher redshift galaxies. They also show that this evolution is highly aperture dependent. Using a fixed physical aperture for all galaxies, in their case 5 kpc, leads to a dark matter fraction that is almost constant over time. Using the stellar half-mass radius instead reveals the evolution of the dark matter fraction with redshift. This is especially evident for galaxies in the $10^{11} \textrm{M}_{\odot}$ stellar mass regime, which are highly concentrated at high redshifts. At a fixed stellar mass these galaxies show a substantial increase in size, leading to smaller half-mass radii at higher redshifts. 

Observationally, various surveys provide some constraints of the dark matter fraction at different redshifts. The DiskMass survey of local galaxies finds that the central dark matter fractions are in the range of 0.5-0.9 within 2.2 times the disc scale radius, which corresponds to $\sim 1.6$ times the half-light radius \citep[][]{Martinsson_2013}. The SWELLS survey \citep[][]{Barnabe_2012, Dutton_2013, Courteau_2015} finds lower dark matter fractions in the range of 0.1 - 0.4 using the same aperture. This discrepancy is most likely due to the SWELLS galaxies having larger bulge components than the DiskMass survey. Galaxies in the redshift range z = 0.6 - 1.2 show a median of $f_{\rm{DM}} \sim 0.3$, while galaxies in the redshift range $z=1.2-2.5$ have a median of $f_{\rm{DM}} \sim 0.12$ within the half-light radius \citep[]{Genzel_2020}. For higher redshifts the dark matter fraction goes as low as $f_\textrm{DM} = 0.05$ \citep[][]{Price_2020} within the half-light radius.

In Q2131-G1 we find a dark matter fraction within the half-light radius of {$f_{\rm{DM}} = 0.24-0.54$}. We therefore find that the central regions of this galaxy are dominated by baryons. Compared to the surveys and simulations, Q2131-G1 fits well between the dark matter fractions found in the DiskMass survey and is also consistent with the median of galaxies observed in the redshift range 0.6-1.2. It is also within the range of galaxies observed within the SWELLS survey. The dark matter fraction is comparable to the one found in galaxies in TNG at the stellar mass $M_{\star} = 10^{10.5} \textrm{M}_{\odot}$ at redshift $z = 2$. While the redshift of these galaxies in TNG is higher than of Q2131-G1, the galaxies within the SWELLS survey fit into the same regime of galaxies found in TNG. Therefore, this discrepancy could partly also be due to Q2131-G1 possibly having a significant bulge component.

\subsection{CO Detection Rate of MUSE-ALMA Halos Survey}

We target the CO(3--2) line of nine galaxies associated with six absorbers with ALMA and detect four of them ($\sim 45$ per cent detection rate). {All of the non-detected galaxies have metallicities below 12 + log(O/H$) \sim 8.65$.} {{Four} of the non-detected galaxies have sub-solar metallicites of {12 = log(O/H) $< 8.32$}, but have higher molecular gas mass limits than the detected galaxy Q2131-G1.} Molecular gas in galaxies with sub-solar gas phase metallicity is shown to be deficient in CO, due to the CO molecule being photo-dissociated at larger fractions compared to higher metallicity galaxies \citep[][]{Wolfire_2010, Bolatto_2013}. This in turn leads to a lower observed CO flux density and longer integration times are needed for observing low metallicity galaxies in CO. We therefore attribute these non-detection to the low metallicites of the galaxies.

\section{Conclusion}

In this paper, we present MUSE and new ALMA observations of the fields Q2131-1207, Q0152-2001, Q0152-2001, Q1211+1030 with LLS, sub-DLAs and DLAs at $z \sim 0.4$ and $z \sim 0.75$. We also include the previously published field Q1130-1449 with 3 CO-detected galaxies ($z \sim 0.3$) in our analysis \citep[][]{Peroux_2019}. We detect one counterpart (Q2131-G1) of a previously detected (HST \& MUSE) galaxy with ALMA observing the CO(3--2) emission line in the field Q2131-1207. We analyse the morphological, kinematical and physical properties of Q2131-G1 with a focus on the molecular gas content. For the non-detections we provide limits on the molecular gas mass and depletion time. 

The findings can be summarised as follows:

\begin{itemize}
     \item The ionised gas phase in Q2131-G1 has a shape indicating spiral arms and possible tidal tails from previous interactions and an extent of $\sim 40$ kpc. The molecular gas is found in a more compact and {elliptical} morphology of smaller extent ($\sim 20$ kpc). The extent of the stellar continuum is in between the ionised and molecular gas phase.
    \item Using the sophisticated 3D forward modeling tool $\textrm{GalPak}^{\textrm{3D}}$ we study the kinematics of the ionised and molecular gas phase of Q2131-G1. We assume a disk model with an exponential flux profile and a tanh rotation curve for both gas phases and find that the gas phases align well directionally with similar inclinations ({$i_\textrm{[OIII]} = (60.5 \pm 1.2) ^{\circ}$}, {$i_\textrm{CO} = (47^{+10}_{-1}) ^{\circ}$})  and position angles ({$PA_\textrm{[OIII]} = (65 \pm 1) ^{\circ}$}, $PA_\textrm{CO} = (59 \pm 2) ^{\circ}$). The maximum rotational velocity is equal for both gas phases ({$V_\textrm{max} \sim 200 \: \textrm{km s}^{-1}$}). This is consistent with findings by the EDGE-CALIFA survey \citep[][]{Levy_2018}, where a fraction of 25 per cent of their sample contained galaxies with equal maximum velocities. We therefore conclude that the ionised and molecular gas phase are strongly coupled within Q2131-G1.
    \item {The absorber shows a neutral and molecular absorption two-component profile, with the weaker component blueshifted and the stronger component redshifted compared to the systemic redshift derived from the kinematic model of the CO emission. Extrapolating the model velocity maps towards the line of sight of the quasar shows that the weaker absorption component is consistent with being part of the extended rotating disk of Q2131-G1. Thanks to metallicity, geometry and orientation arguments, we find that the stronger component is consistent with being gas falling onto Q2131-G1. The considerable amount of molecular gas traced by the absorber poses the question of the presence of a molecular cold phase in infalling gas.}
    \item The molecular mass ({$M_\textrm{mol} = 3.52 ^{+3.95}_{-0.31} \times 10^9 \; \textrm{M}_\odot$}) is on the low end of previously detected HI-selected galaxies. A similar counterpart, associated with the absorber at redshift $z$ = 0.101 in the quasar spectrum of PKS 0429-433 \citep[][]{Neeleman_2016}, interestingly has a similar molecular mass and shows roughly the same HI column density ($\textrm{log}(N_\textrm{HI}/\textrm{cm}^{-2}) \sim 19.5$) and $\textrm{H}_{2}$ column density ($\textrm{log}(N_{\textrm{H}_2}/\textrm{cm}^{-2}) \sim 16.5$). While the absorption was attributed to the CGM of the galaxy and not being part of the rotating disk {or infalling gas}, the similarities of these properties are striking. We conclude that future studies of molecular gas in both absorbers and absorber hosts are essential to studying a possible connection of these properties.
    \item We compute a dark matter fraction within the half-light radius of {$f_{\rm{DM}} = 0.24-0.54$}, showing that the inner parts of the galaxy are baryon dominated. The dark matter fraction fits between the dark matter fraction of the DiskMass survey \citep[$f_\textrm{DM} \sim 0.5-0.9$,][]{Martinsson_2013} and the median dark matter fractions observed in the redshift range $z$ = 0.6-1.2 \citep[$f_\textrm{DM} \sim 0.3$,][]{Genzel_2020}, providing a further indicator for a redshift evolution of the dark matter fraction.
    \item The depletion times (including upper limits) of our sample are in the range of ({$\tau_\textrm{dep} \sim 1.4 - 37 \; \textrm{Gyr}$}). The depletion times of the CO-detected galaxies Q2131-G1, Q1130 and Q1130-G6 are an order of {$\sim 2 - 53$ times} larger than the median depletion time for emission-selected galaxies in the {xCOLD GASS} \citep[][]{Saintonge_2017} and PHIBSS \citep[][]{Tacconi_2018} samples. This result is consistent with previously detected  HI-selected galaxies which also showed higher depletion times compared to emission selected samples. {The high depletion times are a consequence of the high molecular gas masses of HI-selected galaxies for their low SFR. We therefore conclude that HI-selected galaxies possibly preferentially select galaxies that have large molecular gas reservoirs for their low SFR, while a complete picture of the HI-selected population should be obtained by following up the non-detected galaxies for further studies of this possible selection bias.}
    \item The 5 non-detected galaxies all have metallicities below 12 = log(O/H) $\sim$ 8.65. {{Four} of the non-detected galaxies have low sub-solar metallicites of {12 = log(O/H) $< 8.32$}, but have higher molecular gas mass limits than the detected galaxy Q2131-G1.} Combined with the evidence that CO is photo-dissociated at larger fractions in low metallicity galaxies compared to higher metallicity galaxies \citep[][]{Wolfire_2010, Bolatto_2013} and therefore having a lower CO flux density leads to the conclusion that one should account for a higher integration time when observing CO in {sub-solar} metallicity galaxies.
\end{itemize}

\section*{Acknowledgements}

We thank the anonymous referee for the very helpful and detailed comments which helped to improve the final version of the manuscript. We want to thank the ALMA staff for performing the observations. {This paper makes use of the following ALMA data: ADS/JAO.ALMA\#2017.1.00571.S. ALMA is a partnership of ESO (representing its member states), NSF (USA) and NINS (Japan), together with NRC (Canada), MOST and ASIAA (Taiwan), and KASI (Republic of Korea), in cooperation with the Republic of Chile. The Joint ALMA Observatory is operated by ESO, AUI/NRAO and NAOJ.} Based on observations made with the NASA/ESA Hubble Space Telescope, and obtained from the Hubble Legacy Archive, which is a collaboration between the Space Telescope Science Institute (STScI/NASA), the Space Telescope European Coordinating Facility (ST-ECF/ESA) and the Canadian Astronomy Data Centre (CADC/NRC/CSA). We thank Nicolas Bouché developing and distributing the $\textrm{GalPak}^{\textrm{3D}}$ algorithm and for the interactions on the $\textrm{GalPak}^{\textrm{3D}}$ algorithm. R.S. thanks ESO and the IMPRS program for the support of his PhD. A.K. gratefully acknowledges support from the Independent Research Fund Denmark via grant number DFF 8021-00130. VPK gratefully acknowledges support from NASA grant 80NSSC20K0887 and US NSF grant AST/2007538. 

\section*{DATA AVAILABILITY}

Used VLT/MUSE data is available through the ESO Science Archive Facility (http://archive.eso.org/cms.html), ALMA data is available through the ALMA Science Archive (https://almascience.eso.org/asax/) and HST data through the Hubble Legacy Archive (https://hla.stsci.edu). The corresponding IDs are noted in Section \ref{sec:obs}.




\bibliographystyle{mnras}
\bibliography{example} 


\newpage 



\appendix
\begin{landscape}
\section{Observation Details}
\label{sec:alma_obs}

We provide further information about the ALMA observations of the MUSE-ALMA Halos sample used in this work and provide additional information about the QSOs and observed galaxies. The additional information can be seen in table \ref{tbl:quasar_spec}.

\begin{table}
 \begin{tabular}{||c c c c c c c c c c c||} 
 \hline
 \textbf{QSO name} & \textbf{QSO alternative name} & $\textbf{\textbf{RA}}_\textrm{\textbf{QSO}}$ & $\textrm{\textbf{DEC}}_\textrm{\textbf{QSO}}$ & $\textrm{\textbf{z}}_\textrm{\textbf{QSO}} {}^\textrm{\textbf{a}}$ & \textbf{Observation Dates} & $\mathbf{T}_\mathbf{exp}$ & $\boldsymbol{\theta}$ & \textbf{Calibrators} & \textbf{PWV} & \textbf{Ant. Config.} \\ [0.5ex]
 & & & \textbf{[hh:mm:ss]} &  \textbf{[dd:mm:ss] }& & \textbf{[hrs]} & \textbf{['']} & & \textbf{[mm]} \\
 \hline
 Galaxy &
 $\textrm{f}_{\textrm{CO}}$ &
 $\textrm{RA}_\textrm{gal} $&
 $\textrm{DEC}_\textrm{gal} $& 
 $\textrm{z}_\textrm{gal}$ & Detected in CO\\
 & [GHz] & [hh:mm:ss] &  [dd:mm:ss] \\
 \hline\hline
 \textbf{Q2131-1207} & \textbf{Q2128-123} & \textbf{21:31:35} & \textbf{-12:07:04.8} & \textbf{0.43} & \textbf{4, 5, 7 Jun 2018} & \textbf{2.0} & \textbf{1.02} & \textbf{J2148+0657, J2158-1501} & \textbf{0.65 - 2.8} & \textbf{C43-1} \\
 \hline
 Q2131-G1 & {241.866} & 21:31:35.636 & -12:07:00.177 & {0.42974} & yes \\
 \hline
 Q2131-G2 & {241.697} & 21:31:35.775 & -12:07:11.558 &  $0.4307 \;^\textrm{\textbf{a}}$ & no \\
 \hline
 \textbf{Q1232-0224} & \textbf{1229-021} & \textbf{12:32:00} & \textbf{-02:24:04.6} & \textbf{1.05} & \textbf{26, 28 Jun 2018} & \textbf{2.15} & \textbf{1.02} & \textbf{J1218-0119, J1229+0203} & \textbf{1.2 - 2.2} & \textbf{C43-1} \\
 \hline
 Q1232-G1 & 247.829 & 12:31:59.943 & -02:24:05.275 &  $0.3953\;^\textrm{\textbf{a}}$ & no \\
 \hline
 Q1232-G2 & 262.462 & 12:31:59.727 & -02:24:12.20 & $0.7566\;^\textrm{\textbf{a}}$ & no\\
 \hline
 \textbf{Q0152-2001} & \textbf{UM 675} & \textbf{01:52:27} & \textbf{-20:01:07.1} & \textbf{2.06} & \textbf{2, 11, 12 Jul 2018} & \textbf{1.5} & \textbf{0.96} & \textbf{J0006-0623, J0151-1732} & \textbf{0.65 - 2.6} & \textbf{C43-1} \\ 
 \hline
 Q0152-G1 & 250.105 & 01:52:27.827 & -20:01:13.991 &  $0.3826 \;^\textrm{\textbf{a}}$ & no \\
 \hline
 \textbf{Q1211-1030} & \textbf{1209+107} & \textbf{12:11:41} & \textbf{+10:30:02.8} & \textbf{2.19} & \textbf{23 Aug 2018} & \textbf{0.75} & \textbf{0.73} & \textbf{J1229+0203, J1222+0413} & \textbf{0.9 - 1.2} & \textbf{C43-3} \\
 \hline
 Q1211-G1 & {248.274} & 12:11:40.899 & 10:30:06.990 &  $0.3928 \;^\textrm{\textbf{a}}$ & no \\
 \hline
 \textbf{Q1130-1449} & \textbf{1127-145} & \textbf{11:30:07} & \textbf{-14:49:27.7} & \textbf{1.19} & \textbf{4, 8, 15 Dec 2016} & \textbf{3.6} & \textbf{1.13} & \textbf{J1058+0133, J1139-1350} & \textbf{1.5 - 5.4} & \textbf{C40-3} \\
 \hline
  Q1130-G2 & 263.4 & 11:30:07.66 & -14:49:23.41 &  $0.3127 \;^\textrm{\textbf{a}}$ & yes \\
 \hline
  Q1130-G4 & 263.44 & 11:30:07.62 & -14:49:11.44 &  $0.3126 \;^\textrm{\textbf{a}}$& yes \\
 \hline
   Q1130-G6 & 263.67 & 11:30:08.53 & -14:49:28.54 & $0.3115\;^\textrm{\textbf{a}}$ & yes \\
 \hline
 \hline

\end{tabular}
\caption{\textbf{\label{tbl:quasar_spec} Properties of the quasars and galaxies in the MUSE-ALMA Halo sample.} \newline
Row 1: (1) reference name of the QSO used in this paper (2) full name of the QSO, (3) right ascension of the QSO, (4) declination of the QSO, (5) QSO redshift, (6) dates for the observations of the field, (7) exposure time of the observation, (8) angular resolution of the observation, (8) calibrators used for the observation, (9) percipitable water vapour (PWV) of the observation, (9) ALMA antenna configuration used for the observation \newline
Row 2: (1) reference name of the galaxy used in this paper, (2) redshifted frequency of the observed CO line, (3) right ascension of the galaxy, (4) declination of the galaxy, (5) redshift of the galaxy, (5)  \newline
Literature references:  ${}^\textrm{\textbf{a}}$) \protect\cite{Hamanowicz_2020}}
\end{table}

\end{landscape}

\section{Q2131-G1 - Kinematic Modelling Residuals and Model Flux Map}

\begin{figure}[hbt!]
  \captionsetup{}
  \includegraphics[width=1.0\textwidth]{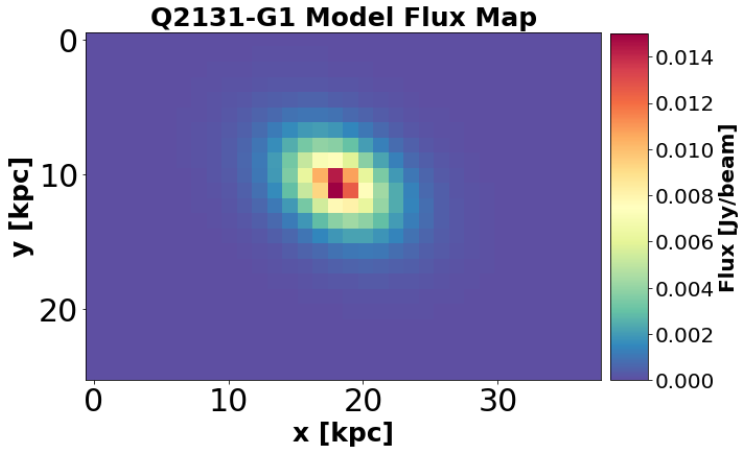}
  \caption{{Flux map of Q2131-G1 modelled in 3D-space with $\textrm{GalPak}^{\textrm{3D}}$.}}
  \label{fig:2131G1_mod_fl}
\end{figure}

\begin{figure}[hbt!]
  \captionsetup{}
  \includegraphics[width=1.0\textwidth]{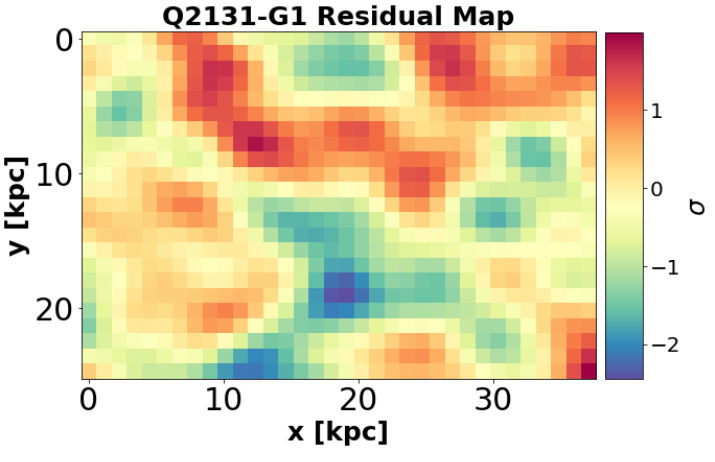}
  \caption{{Residual flux map of Q2131-G1 between the modelled and observed fluxes. The low residuals show that the disk model reproduces the observations well.} {The colorbar displays data - model normalized by the pixel noise $\sigma$.}}
  \label{fig:2131G1_residual}
\end{figure}

The model {molecular gas} flux map and residuals of the galaxy Q2131-G1 derived from $\textrm{GalPak}^{\textrm{3D}}$ can be seen in Figures \ref{fig:2131G1_mod_fl} and \ref{fig:2131G1_residual}. The modelled disk reproduces the observations well, as can be seen by the low residuals.

\newpage
\newpage
\begin{figure}[hbt!]
  \captionsetup{}
  \includegraphics[width=1.0\textwidth]{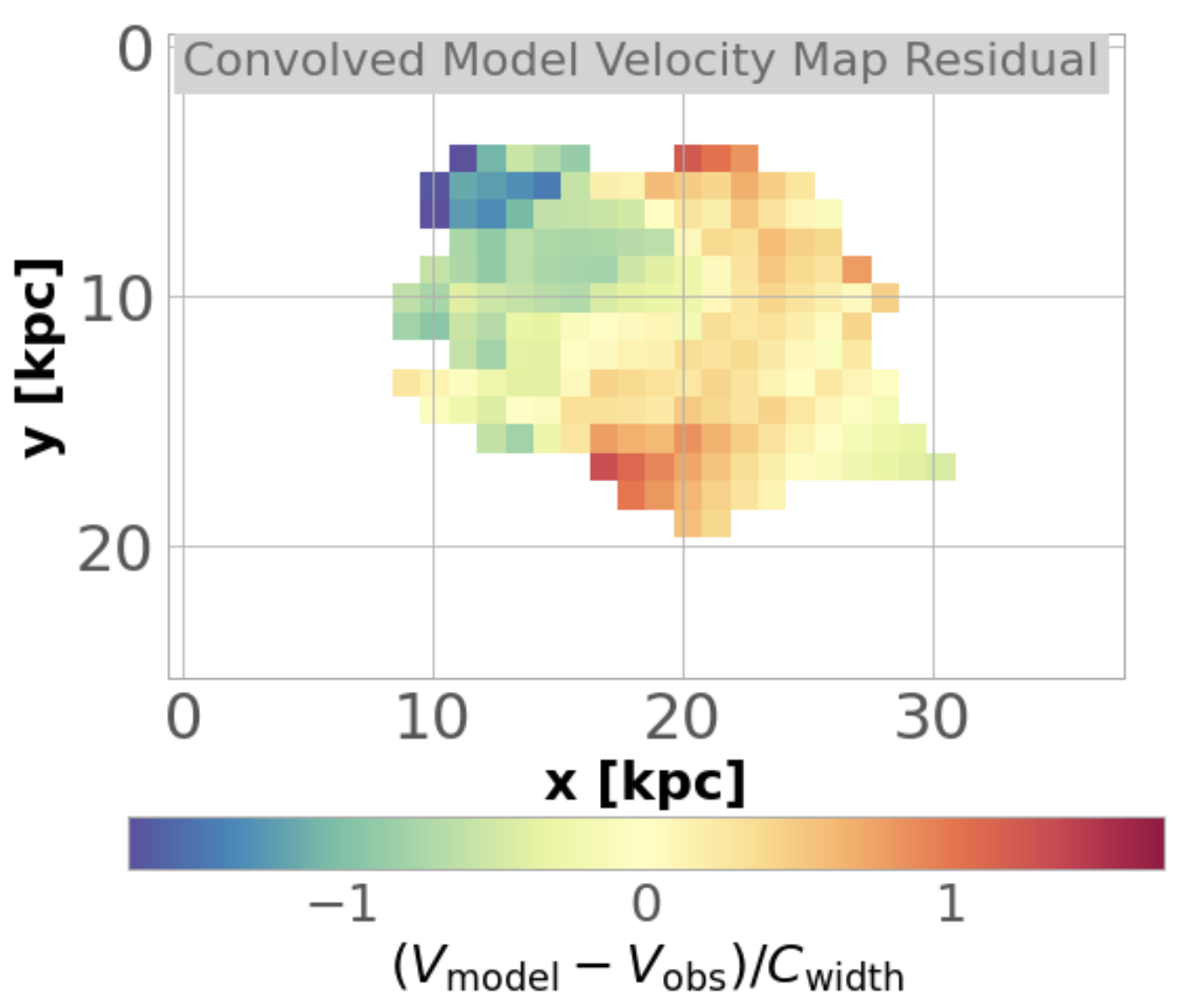}
  \caption{{Convolved Model Velocity Map Residual. The colorbar is the residual divided by the spectral resolution of the cube ($C_\textrm{width} = 50 \textrm{km s}^{-1})$. The low residuals across the galaxy indicate that the disk model with an arctan velocity profile reproduces the observations robustly.}}
  \label{fig:2131G1_vel_res}
\end{figure}


\bsp	
\label{lastpage}
\end{document}